\documentclass[12pt]{article}
\pdfoutput=1
\usepackage[a4paper]{geometry}
\usepackage{jheppub, amsmath,amssymb,amsfonts,amsxtra, mathrsfs, makeidx,graphics,graphicx,amsthm,epsfig, ytableau,bm,longtable,float, color,tikz,mathtools,xfrac,footnote,rotating, lscape, makecell, environ,mathtools, empheq}

\usetikzlibrary{positioning}
\usetikzlibrary{chains}
\usetikzlibrary{arrows,fit,decorations.pathreplacing}
\tikzstyle{every picture}+=[remember picture]
\tikzstyle{na} = [baseline]

\usetikzlibrary{arrows, decorations.markings, calc, fadings, decorations.pathreplacing, patterns, decorations.pathmorphing, positioning}

\def\node#1#2{\overset{#1}{\underset{#2}{{\color{gray} \bullet}}}}

\def\sqnode#1#2{\overset{#1}{\underset{#2}{{\color{gray} \blacksquare}}}}

\def\sqwnode#1#2{\overset{#1}{\underset{#2}{{\square}}}}

\def\cvver#1#2{\overset{{\llap{$\scriptstyle#1$}\displaystyle{\color{gray} \bullet}{\rlap{$\scriptstyle#2$}}}}{\scriptstyle\uparrow\, \downarrow}}

\def\bnode#1#2{\overset{#1}{\underset{#2}{{\color{blue} \bullet}}}}

\def\bsqnode#1#2{\overset{#1}{\underset{#2}{{\color{blue} \blacksquare}}}}

\def\bsqcver#1#2{\overset{{\llap{$\scriptstyle#1$}\displaystyle{\color{blue} \blacksquare}{\rlap{$\scriptstyle#2$}}}}{\scriptstyle\vert}}

\def\node#1#2{\overset{#1}{\underset{#2}{\circ}}}
\def\sqnode#1#2{\overset{#1}{\underset{#2}{{\color{gray} \blacksquare}}}}
\def\rnode#1#2{\overset{#1}{\underset{#2}{{\color{red} \bullet}}}}
\def\bnode#1#2{\overset{#1}{\underset{#2}{{\color{blue} \bullet}}}}
\def\gnode#1#2{\overset{#1}{\underset{#2}{{\color{gray} \bullet}}}}
\def\sqgrnode#1#2{\overset{#1}{\underset{#2}{{\color{gray} \blacksquare}}}}
\def\sqbnode#1#2{\overset{#1}{\underset{#2}{{\color{blue} \blacksquare}}}}
\def\sqrnode#1#2{\overset{#1}{\underset{#2}{{\color{red} \blacksquare}}}}

\def\sqwnode#1#2{\overset{#1}{\underset{#2}{{ \square}}}}

\def\ver#1#2{\overset{{\llap{$\scriptstyle#1$}\displaystyle\circ{\rlap{$\scriptstyle#2$}}}}{\scriptstyle\vert}}
\def\wver#1#2{\overset{{\llap{$\scriptstyle#1$}\displaystyle{\square}{\rlap{$\scriptstyle#2$}}}}{\scriptstyle\vert}}
\def\bluesqver#1#2{\overset{{\llap{$\scriptstyle#1$}\displaystyle{\color{blue} \blacksquare}{\rlap{$\scriptstyle#2$}}}}{\scriptstyle\vert}}

\def\grvcver#1#2{\overset{{\llap{$\scriptstyle#1$}\displaystyle{\color{gray} \bullet}{\rlap{$\scriptstyle#2$}}}}{\scriptstyle\uparrow\, \downarrow}}
\def\grvver#1#2{\overset{{\llap{$\scriptstyle#1$}\displaystyle{\color{gray} \blacksquare}{\rlap{$\scriptstyle#2$}}}}{\scriptstyle\uparrow\, \downarrow}}

\tikzstyle{every picture}+=[remember picture]
\tikzstyle{na} = [baseline=-.5ex]

\usepackage{bbm}
\usepackage{subfigure}

\newcommand{\eg}{\textit{e.g.}}

\newcommand{\ie}{\textit{i.e.}}

\numberwithin{equation}{section}

\newcommand{\nn}{\nonumber}

\newcommand{\be}{\begin{equation}} \newcommand{\ee}{\end{equation}}
\newcommand{\bea}{\begin{equation} \begin{aligned}} \newcommand{\eea}{\end{aligned} \end{equation}}

\def\tilde{\widetilde}

\def\rt2{\sqrt{2}}

\def\det{\mathop{\rm det}}

\def\Tr{\mathop{\rm Tr}}
\def\tr{\mathop{\rm tr}}

\def\CN{{\cal N}}

\def\CX{{\cal X}}

\def\1{{\ds 1}}

\def\repa{\raise4pt\hbox{$\square$}\mkern-14mu\raise-4pt\hbox{$\square$}}
\def\repab{\overline{\raise4pt\hbox{$\square$}\mkern-14mu\raise-4pt\hbox{$\square$}\mkern-1mu}}

\def\smileface{\ensuremath{\hbox{\large$\bigcirc$}\mkern-15mu\raise-1pt\hbox{\scriptsize$\smallsmile$}%
\mkern-10mu\raise4pt\hbox{..}\mkern4mu}}
\def\frownface{\ensuremath{\hbox{\large$\bigcirc$}\mkern-15mu\raise-1pt\hbox{\scriptsize$\smallfrown$}%
\mkern-10mu\raise4pt\hbox{..}\mkern4mu}}

\newcommand{\ba}{\begin{array}}
\newcommand{\ea}{\end{array}}
\newcommand{\bi}{\begin{itemize}}
\newcommand{\ei}{\end{itemize}}
\def\vec#1{\bm{#1}}
\def\bea#1\eea{\allowdisplaybreaks \begin{align}#1\end{align}}
 \newcommand{\ben}{\begin{enumerate}}
\newcommand{\een}{\end{enumerate}}
\newcommand{\bean}{\begin{eqnarray*}}
\newcommand{\eean}{\end{eqnarray*}}
\newcommand{\eref}[1]{(\ref{#1})}

\newcommand{\tQ}{\widetilde{Q}}

\newcommand{\BZ}{\mathbb{Z}}

\newcommand{\W}{\mathcal{W}}

\newcommand{\comment}[1]{}
\newcommand{\diag}{\mathrm{diag}}

\definecolor{light-gray}{gray}{0.7}

\def\arr#1#2{\underset{#2}{\overset{#1}{\substack{\longrightarrow \\ \longleftarrow}}}}
\def\alr#1{\,\, \underset{\tilde{#1}}{\overset{#1}{\substack{\longrightarrow \\ \longleftarrow}}} \,\,\,}
\def\aup#1 {\overset{#1}{\uparrow} \, \overset{\tilde{#1}}{\downarrow}}

\title{Mirror theories of 3d $\CN=2$ SQCD}
\author[a,b]{Simone Giacomelli}
\author[c,d]{and Noppadol Mekareeya}

\affiliation[a]{International Center for Theoretical Physics, \\ Strada Costiera 11, 34151 Trieste, Italy}
\affiliation[b]{INFN, Sezione di Trieste, Via Valerio 2, 34127 Trieste, Italy}
\affiliation[c]{INFN, Sezione di Milano-Bicocca, Piazza della Scienza 3, I-20126 Milano, Italy}
\affiliation[d]{Dipartimento di Fisica, Universit\`a di Milano-Bicocca, \\ Piazza della Scienza 3, I-20126 Milano, Italy}
\emailAdd{sgiacome@ictp.it}
\emailAdd{n.mekareeya@gmail.com}

\abstract{Using a recently proposed duality for $U(N)$ supersymmetric QCD (SQCD) in three dimensions with monopole superpotential, in this paper we derive the mirror dual description of $\mathcal{N}=2$ SQCD with unitary gauge group, generalizing the known mirror dual description of abelian gauge theories. We match the chiral ring of the dual theories and their partition functions on the squashed sphere. We also conjecture a generalization for SQCD with orthogonal and symplectic gauge groups.}

\begin{document}
\maketitle

\section{Introduction} 
A remarkable feature of supersymmetric gauge theories is the existence of infrared dualities: two seemingly different gauge theories become equivalent at low energies. One of the most important properties of these correspondences is the fact that quantities which are hard to compute in one thery due to nonperturbative effects are often mapped to easier problems in the dual description. One well-known example is the structure of the Coulomb branch of three-dimensional $\CN=4$ theories, which is subject to quantum corrections. Using mirror symmetry \cite{Intriligator:1996ex}, one can argue that this is equivalent to the Higgs branch of the mirror theory, which instead can be reliably studied using the classical equations of motion due to a nonrenormalization theorem. 

By now we have many examples of this phenomenon and this led to many new insights about the dynamics of supersymmetric gauge theories. On the other hand, at present we do not have a systematic understanding of infrared dualities and an algorithm to extract them is not available (yet). Ideally, we may wish to have the following result: starting from a small set of prototypical examples, such as Seiberg duality in four dimensions \cite{Seiberg:1994pq} or mirror symmetry for three-dimensional theories with eight supercharges \cite{Intriligator:1996ex}, one is allowed to modify the matter content and superpotential interactions of the theory by applying a ``canonical'' set of operations. If on top of this we are able to map these operations on the dual side, then we can systematically extract dual descriptions for other gauge theories. 

The purpose of this paper is to make some progress in this direction in the context of mirror symmetry in three dimensions: as is well known, the mirror map is understood for a very large class of theories, especially those with eight supercharges. This was achieved with a variety of arguments including stringy-inspired constructions \cite{Hanany:1996ie, deBoer:1996mp, Porrati:1996xi, deBoer:1996ck, deBoer:1997ka, Assel:2014awa, Benvenuti:2016wet}. One natural question is then whether this family of dualities can be extended to more general 3d $\CN=2$ theories. This is rather well understood in the case of abelian theories, since the required modification of the matter content is rather easy to implement: in the $\CN=2$ language a $\CN=4$ vector multiplet includes a chiral multiplet $\Phi$ in the adjoint representation of the gauge group (hence we are dealing with gauge singlets in the abelian case) and extended supersymmetry implies the presence of cubic suerpotential terms involving these chiral multiplets in the adjoint. In order to derive a mirror dual for the pure $\CN=2$ abelian theory (see \cite[sec. 4]{Aharony:1997bx}), it is enough to introduce by hand a gauge singlet $S$ and turn on the superpotential term $S\Phi$. This makes both singlets massive and removes all cubic superpotential terms, so at low energy we are left with the pure $\CN=2$ theory. This procedure can be implemented on the mirror side as-well: since in the abelian case $\Phi$ is a gauge invariant chiral operator, it should have a counterpart in the mirror description so it is enough to add by hand a singlet $S'$ in the mirror theory and couple it to the mirror image of $\Phi$. 

This construction does not extend to the nonabelian case since in this case $\Phi$ is no longer gauge invariant and it is not obvious how introducing a second chiral multiplet in the adjoint representation affects the dual theory. This is precisely the problem we will discuss in the present note. Our basic observation is that the $\CN=4$ linear quiver 
\be \label{TSUN}
\node{}{1}-\node{}{2}-\cdots- \node{}{N-1}-\sqwnode{}{N}~,
\ee
usually called $T(SU(N))$ in the literature \cite{Gaiotto:2008ak}, flows in the IR to a free theory consisting of a chiral multiplet in the adjoint of the $SU(N)$ global symmetry upon a certain monopole superpotential deformation we will describe in detail. The idea is then the following: in order to extract the mirror dual of a $\CN=2$ $SU(N)$ theory with zero superpotential, we start from its $\CN=4$ counterpart and we couple to it $T(SU(N))$. In many cases the mirror of this $\CN=4$ theory can be extracted using the methods already available in the literature (see \eg~ \cite{Aharony:1997bx, Gaiotto:2008ak, Assel:2011xz, Cremonesi:2014kwa}). Then we activate the suitable monopole superpotential for $T(SU(N))$, which reduces (due to our observation) to a chiral multiplet in the adjoint of the now gauged $SU(N)$ symmetry. The ordinary $\CN=4$ superpotential coupling reduces to a quadratic term which makes both the adjoint in the $\CN=4$ vector multiplet and the newly-created adjoint massive, so the theory becomes equivalent at low energy to a pure $\CN=2$ theory. The monopole superpotential is mapped on the mirror side to superpotential terms involving the off-diagonal components of the meson (or more precisely the $SU(N)$ moment map) so in this way we extract the candidate mirror dual for the $\CN=2$ theory. 

In principle this procedure can be repeated multiple times, allowing to vary at will the number of adjoint chiral multiplets in the theory. We will see that for $USp(2N)$ gauge theories this procedure allows to vary both the number of adjoints and also the number of traceless antisymmetric chiral multiplets. The main issue is that, when this procedure is used to introduce new matter fields, the theory frequently exhibits emergent symmetries in the infrared which do mix with the $R$-symmetry and these are not manifest in the dual description. One should then also understand how to detect them in order to extract information about the infrared fixed point. 

The paper is organized as follows. In Section \ref{tsusect} we show that upon a suitable monopole superpotential deformation $T(SU(N))$ reduces to a chiral multiplet in the adjoint representation of $SU(N)$. We first present a field-theoretic argument using a recently discovered duality for $U(N)$ SQCD with monopole superpotential and then match partition functions on the squashed sphere. In Section \ref{su2mirr2} we use this observation to extract the mirror dual of $SU(2)$ SQCD. Since in this case the dual model is relatively simple, we can perform a detailed match of the chiral ring of the dual theories. In Section \ref{suNmirr2} we generalize the result to SQCD with gauge group $U(N)$ and $SU(N)$. We also discuss the matching of squashed-sphere partition functions. In Section \ref{sec:branes} we discuss the brane interpretation of our results and in Section \ref{antiusp} we apply the same idea to extract the mirror dual of $USp(2N)$ SQCD with fundamental and antisymmetric matter. In Appendix \ref{sec:orthosymp} we provide a proposal for the mirror dual of SQCD with orthogonal and symplectic gauge groups, finding nontrivial agreement at the level of the chiral ring. The derivation in this case would require the generalization of the arguments presented in Section \ref{tsusect} to $T(SO(2N))$ theory.

\section{Monopole superpotentials and confinement}\label{tsusect} 
\subsection{$U(N)$ SQCD and the monopole duality}

The main tool used in this paper is the monopole duality found in \cite{Benini:2017dud}: the following gauge theories 
\begin{itemize} 
 \item {\bf Theory A.} $U(N_c)$ SQCD with $N_f$ flavors and monopole superpotential $\mathcal{W}=V^+$ (where $V^+$ of course denotes the monopole operator with magnetic flux +1), 
 \item {\bf Theory B.} $U(N_f-N_c-1)$ SQCD with $N_f$ flavors and superpotential $\mathcal{W}=M_{ij}\widetilde{Q}^iQ^j+V^-+\CX V^+$, where $M_{ij}$ and $\CX$ are gauge singlets 
\end{itemize}
flow to the same IR fixed point. This is derived by reducing to 3d the 4d Intriligator-Pouliot duality for $Usp(2N_c)$ SQCD and turning on real masses to break the gauge group to $U(N_c)$. We will be primarily interested in the special case $N_f=N_c+1$, in which theory B reduces to a Wess-Zumino model and the duality becomes 
\be \label{dualconfine}
\begin{split}
&\text{$U(N_c)$ with $N_f=N_c+1$ with $W=V^+$} \\
&\longleftrightarrow \quad \text{$N_f^2$ singlets $M$ and a singlet $\gamma$} \\
& \qquad \qquad \text{with $\W= \gamma\det(M)$}~,
\end{split}
\ee
where $\gamma$ is dual to the monopole $V^-$ in theory A and $M$ is the counterpart of the meson $\widetilde{Q}^iQ^j$ in theory A. For $N_c=1$ (\ref{dualconfine}) can also be extracted from mirror symmetry (see \cite{Collinucci:2016hpz}). We will now see that by turning on a suitable monopole superpotential and repeatedly using (\ref{dualconfine}), $T(SU(N))$ can be converted into a single chiral multiplet in the adjoint of $SU(N)$. Our construction is essentially a variant of the method described in \cite{Benvenuti:2017kud}.

\subsection{Monopole deformation of $T(SU(N))$} 

Let us start from the simplest case, namely $T(SU(2))$ which is just $\CN=4$ SQED with two flavors. We now introduce a singlet $X$ and turn on two monopole superpotential terms: $\delta\W= V^+ + \CX V^-$. The full superpotential of the theory is now (we denote with $\phi$ the chiral mutiplet in the $\CN=4$ vector multiplet) 
\be\W=\phi\widetilde{Q}_iQ^i+V^+ + \CX V^-\ee 
Using now (\ref{dualconfine}) we conclude that this theory is equivalent to a WZ model with superpotential 
\be\W=\gamma\det(M)+\phi\tr(M)+\CX \gamma.\ee 
We immediately see that $\phi$, $\gamma$, $\CX$ and $\tr(M)$ are massive and at low energy we are left with the traceless part of $M$ (i.e. an adjoint of $SU(2)$) and zero superpotential. This is precisely the claim made above. 

The idea for the general case is simply to iterate the above steps. In order to understand how this works, let us discuss $T(SU(3))$, which is the following $\CN=4$ linear quiver with two gauge nodes: 
\be \node{}{1}- \node{}{2} - \sqwnode{}{3}~. \ee
In $\CN=2$ notation, the above quiver can be written as
\be
{\gnode{\overset{\phi_1}{\cap}}{1}} \alr{Q} {\gnode{\overset{\phi_2}{\cap}}{2}} \alr{P} \sqgrnode{}{3}~.
\ee
We denote with $\phi_1$ and $\phi_2$ the adjoint chirals in the $U(1)$ and $U(2)$ vector multiplets respectively. We denote the $U(1)\times U(2)$ bifundamental hypermultiplet with $Q$ and $\widetilde{Q}$ and the three $U(2)$ doublets with $P_i$ and $\widetilde{P}_i$. We denote with $V^{a,b}$ the monopoles with magnetic flux $a$ relative to the $U(1)$ gauge group and magnetic flux $(b,0)$ under the $U(2)$ group. The superpotential of the theory is 
\be\W=\phi_1\widetilde{Q}Q+\tr[\phi_2(\widetilde{P}_iP^i-Q\widetilde{Q})].\ee 
As in the previous case, we turn on superpotential terms involving the monopoles charged under the $U(1)$ group: $\delta\W=V^{+0}+\CX_1V^{-0}$. From (\ref{dualconfine}) we conclude that the $U(1)$ node confines and is traded for an adjoint of $SU(2)$. The resulting theory is $U(2)$ SQCD with 3 flavors, two chirals in the adjoint and superpotential 
\be\label{step1}\W=\CX_1\gamma+\gamma\det(M)+\phi_1\tr M+\tr[\phi_2(P_i\widetilde{P}^i-M)].\ee 
Because of the mass terms both adjoints can be integrated out and we are left with $U(2)$ SQCD with three flavors and the singlet $\tr\phi_2$. The superpotential is simply 
\be\label{step2}\W=\tr\phi_2\tr\widetilde{P}_iP^i.\ee 
Since this theory has no adjoints, we are in the position to apply (\ref{dualconfine}) again, provided we add the superpotential terms $\delta\W=W^+ + \CX_2W^-$ (where $W^{\pm}$ are the $U(2)$ monopoles with topological charge $\pm1$). Once this deformation is turned on, the $U(2)$ group confines and we are left with an adjoint of $SU(3)$ (the trace part becomes massive due to (\ref{step2})) and zero superpotential as desired. 

Our goal is then to find the proper monopole superpotential which reduces, once the $U(1)$ group is confined, to $W^+ + \CX_2W^-$. A very similar setup was already considered in \cite{Benvenuti:2017kud}, where it was observed that $V^{0+}$ is mapped to $W^+$ after confinement of the $U(1)$ and analogously $V^{--}$ is mapped to $W^-$. This prompts us to turn on the superpotential terms $V^{0+}+\CX_2V^{--}$. $V^{0-}$ instead becomes equivalent in the chiral ring to $V^{-0}$ (or more precisely $\gamma$ appearing in (\ref{step1})) once the $U(2)$ node as well is confined (see the discussion around \cite[Eq. (3.9)]{Benvenuti:2017kud}). In conclusion, our prescription is to deform $T(SU(3))$ by turning on the superpotential 
\be\label{monsu3}\delta\W=V^{+0}+V^{0+}+\CX_1(V^{-0}+V^{0-})+\CX_2 V^{--}.\ee

At this stage it should be clear how to proceed in general: we deform $T(SU(N))$ by adding singlets $\CX_1,\dots, \CX_{N-1}$ and turning on the following superpotential 
\be \label{WdefA1}
\begin{split}
\delta\W &=(V^{+00\cdots0}+ V^{0+0\cdots0}+ V^{00+\cdots0} + \ldots  + V^{000\cdots+})  \\
&\qquad + \CX_{1} [ V^{-00\cdots0}+ V^{0-0\cdots0}+ V^{00-\cdots0}+ \ldots (\text{terms with one minus})  ] \\
&\qquad +\CX_{2} [ V^{--0\cdots0}+ V^{0--\cdots0} +\ldots (\text{terms with two minuses}) ] + \ldots \\
&\qquad + \CX_{N-1} V^{---\cdots-}~,
\end{split}
\ee
where $V^{j_1 j_2 j_3 \cdots j_{N-1}}$ are (the notation is the same as before) the monopole operators carrying flux $(j_1, (j_2,0), \ldots, (j_{N-1},\dots,0))$ under $U(1)$, $U(2)$, $\cdots$, $U(N-1)$ gauge groups. Repeatedly applying the monopole duality (\ref{dualconfine}) and integrating out massive fields, we conclude that all the gauge nodes confine and the $SU(N)$ moment map turns into a free chiral multiplet in the adjoint of $SU(N)$. This observation constitutes the main tool of the present paper. 

\subsubsection{The mirror dual of monopole deformed $T(SU(N))$} 

It is instructive to analyze the mirror dual of the superpotential deformation (\ref{WdefA1}) to get a better insight into our procedure. As is well known, $T(SU(N))$ is self-mirror and the monopole operators appearing in (\ref{WdefA1}) are mapped to components of the Higgs branch $SU(N)$ moment map. As a result, the superpotential deformation (\ref{WdefA1}) is equivalent to introducing a field-dependent mass matrix (which depends on the singlets $\CX_i$) of the form: 
\begin{equation}\label{massmat}
M=\left(\begin{array}{ccccc}
0 & 1 & 0 & \dots & 0 \\
\CX_{1} & 0 & 1 & \hphantom{X_{N}} & 0\\
\CX_{2} & \CX_{1} & \ddots & \ddots & \hphantom{X_{N}}\\
\vdots & \ddots & \ddots & 0 & 1\\
\CX_{N-1} & \dots & \CX_{2} & \CX_{1} & 0
\end{array}\right)\ .
 \end{equation} 
We can now make the following observation: introducing a field-dependent mass of this type is equivalent to coupling to the moment map a chiral multiplet in the adjoint of $SU(N)$ and turning on a principal nilpotent vev for it. As a result, all the flavors become massive except one (which we call $q$, $\tilde{q}$) and integrating out massive fields we are left with (see the Appendix A of \cite{Benvenuti:2017kud}) 
\be\label{massdef}\W=\tilde{q}\phi^Nq+\sum_{i=1}^{N-1} \CX_i\tilde{q}\phi^{N-i-1}q,\ee
where $\phi$ is the chiral multiplet in the $\CN=4$ $U(N-1)$ vector multiplet. We shall discuss further details regarding the first term of this superpotential in Section \ref{suNmirr2} and in Appendix \ref{sec:chiralringstab}.  This type of superpotential will appear several times below.

\subsection{The $S^3_b$ partition function} 

The purpose of this section is to test our dual description of $T(SU(N))$ at the level of squashed sphere partition function.
Our conventions are as follows: the contribution of each chiral is \cite{Hama:2010av,Hama:2011ea}  
\be\mathcal{Z}_{\chi}=s_b\left(i\frac{Q}{2}-\tilde{m}_{\chi}\right),\ee 
where $s_b(x)$ is the double sine function ($b$ denotes the squashing parameter):
\be s_b(x)\equiv\prod_{n,m\in\mathbb{Z}\geq0}\frac{bm+nb^{-1}+Q/2-ix}{bm+nb^{-1}+Q/2+ix};\quad Q\equiv b+\frac{1}{b}.\ee
$\tilde{m}_{\chi}$ denotes the following quantity: for every $U(1)$ symmetry $R_i$ we can turn on a real mass $m_i$ and consider its mixing with the $R$-symmetry $R=R_0+\sum_ic_iR_i$. Here $R_0$ denotes some $R$-symmetry and $c_i$ is the mixing coefficient. An important observation is that the partition function on the squashed sphere is holomorphic in $m_i+i\frac{Q}{2}c_i$ for every $U(1)$ symmetry including topological symmetries (in the latter case the real mass is identified with the FI parameters $\xi$) \cite{Jafferis:2010un}. We then define 
\be\label{compmass}\tilde{m}_{\chi}\equiv \sum_iq^i_{\chi}\left(m_i+i\frac{Q}{2}c_i\right),\ee 
where $q^i_{\chi}$ denotes the charge of the chiral multiplet under $R_i$. Notice that $m_0=0$ (there is no real mass relative to the $R$-symmetry $R_0$) and $c_0=1$.

Using this notation the partition function of $T(SU(2))$, i.e. SQED with two flavors, can be written as follows: 
\be
\mathcal{Z}= s_b(m_A)\int_{-\infty}^{\infty} du e^{2\pi iu\xi}
s_b\left(i\frac{Q}{4}-\frac{m_A}{2}+u\pm\frac{m_F}{2}\right)s_b\left(i\frac{Q}{4}-\frac{m_A}{2}-u\mp \frac{m_F}{2}\right),\ee
where $\xi$ denotes the FI parameter, $m_F$ is the fugacity for the $SU(2)$ symmetry acting on the two flavors and $m_A$ is the real mass associated with the $U(1)$ ``axial'' symmetry $H-C$ ($C$ and $H$ denote respectively the Cartan generators of the $SU(2)_C\times SU(2)_H$ $R$-symmetry of the $\CN=4$ theory). This real mass term breaks $SO(4)_R$, hence extended supersymmetry and is usually neglected in writing down the partition function of a theory with eight supercharges and actually several simplifications occur if we set $m_A=0$. However, this parameter will play an important role in the present paper so we prefer keeping it from the start. The partition function of $T(SU(N))$ can then be written recursively as follows: 
\begin{eqnarray}\label{pftsun}
\mathcal{Z}_{T(SU(N))}&=&\frac{1}{(N-1)!}\int\prod_{i=1}^{N-1}du_ie^{2\pi i\xi_{N-1}(\sum_iu_i)}\mathcal{Z}_{T(SU(N-1))}(u_i,\xi_i,m_A)\times \\ 
&& \frac{\prod_{i,j=1}^{N-1} s_b(u_i-u_j+m_A)\prod_{i=1}^{N-1}\prod_{j=1}^{N}s_b\left(i\frac{Q}{4}\pm u_i\mp m_j-\frac{m_A}{2}\right)}{\prod_{i<j}^{N-1}s_b\left(i\frac{Q}{2}\pm(u_i-u_j)\right)}~,\nonumber
\end{eqnarray} 
where the factor $\prod_{i<j}^{N-1}s_b\left(i\frac{Q}{2}\pm(u_i-u_j)\right)$ denotes the contribution from the $U(N-1)$ gauge group, the factor $\prod_{i,j=1}^{N-1} s_b(u_i-u_j+m_A)$ denotes the contribution from the adjoint chiral field under the $U(N-1)$ gauge group, and $\prod_{j=1}^{N}s_b\left(i\frac{Q}{4}\pm u_i\mp m_j-\frac{m_A}{2}\right)$ denotes the contribution from the bifundamental hypermultiplet between the $U(N-1)$ gauge group and the $U(N)$ flavour symmetry.
In the above formula $\xi_{N-1}$ denotes the FI parameter of the $U(N-1)$ gauge symmetry, the parameters $m_j$ (subject to the constraint $\sum_jm_j=0$) are the $SU(N)$ real masses and $m_A$ is again the real mass for the ``axial'' $U(1)$ symmetry described before. The parameters $\xi_i$ ($i=1,\dots,N-2$) denote instead the FI parameters of the gauge groups inside $T(SU(N-1))$. All these parameters can be complexified and the imaginary part describes the mixing with the $R$-symmetry. 

In order to write down the partition function of the monopole deformed $T(SU(N))$ theory, we need first of all to identify the $R$-symmetry of the theory. The effect of the monopole superpotential is to break $\CN=4$ supersymmetry to $\CN=2$ and to mix the $R$-symmetry with the topological symmetries $T_i$ of the theory: our monopole deformation breaks completely the $SU(N)$ Coulomb branch symmetry and the corresponding $N-1$ Cartan generators mix with the $R$-symmetry. The mixing coefficients are determined demanding that the monopole operators $V^{+0\dots0}...$ appearing in (\ref{WdefA1}) have $R$-charge 2. The monopole operator with magnetic flux $(1,0\dots,0)$ under $U(k)$ and trivial flux under all other gauge groups has charge one under $T_k$ and zero charge under all other topological symmetries. Apart from the $T_i$'s, we have to take into account the two $U(1)$ symmetries $C$ and $H$. Our trial $R$-symmetry can be parametrized as follows: 
\be\label{trial} R_{\alpha}=C+H+\alpha(C-H)+(1-\alpha)\sum_iT_i.\ee 
Under this combination,
\bi
\item the adjoint chirals in the $\CN=4$ vector multiplets have $C=1, \, H=0,\, T_i=0$ and thus have charge $R_{\alpha} = 1+\alpha$, 
\item  the bifundamental hypermultiplets have $C=0, \, H=\frac{1}{2},\, T_i=0$ and thus have charge $R_{\alpha} = \frac{1-\alpha}{2}$; and 
\item the monopole operators with charge $+1$ under one $T_i$ generator, \ie~ those appearing in the first line of (\ref{WdefA1}),  have $C=1$, $H=0$; hence they carry charge $R_{\alpha}= 2$. 
\ei
As a result, all superpotential terms in \eref{WdefA1} have $R$-charge exactly 2 provided we assign charge $(i+1)(1-\alpha)$ to the singlets $\CX_i$. The parameter $\alpha$ cannot be determined with these considerations alone and we need to perform $Z$-extremization in order to fix the $R$-symmetry \cite{Jafferis:2010un}. In the rest of this section we will work in terms of the trial $R$-symmetry $R_{\alpha}$\footnote{We would like to notice that in the $\CN=4$ theory without the superpotential (\ref{WdefA1}), the mixing with the topological symmetries can be discarded and we are left with $R_{\alpha}=C+H+\alpha(C-H)$. The result of Z-extremization is $\alpha=0$ for all good or ugly theories.}. Notice that, since the superpotential (\ref{WdefA1}) breaks all the topological symmetries and $C-H$ except the combination $C-H-\sum_iT_i$, all the FI parameters and the real mass for $H-C$ are identified. Throughout this section we will call the resulting parameter $\xi$.

The strategy is to prove our claim by induction: we first check the claim is true for $N=2$ and then show that it holds for $T(SU(N+1))$ assuming it holds for $T(SU(N))$. Let us start by analyzing the $T(SU(2))$ case: the theory is simply SQED with two flavors and monopole superpotential $V^++\CX_1V^-$. The singlet $\CX_1$ has charge $2-2\alpha$ under (\ref{trial}). The partition function then reads: 
\bean
\mathcal{Z} &=& s_b\left(iQ\alpha-i\frac{Q}{2}-2\xi\right)s_b\left(\xi-i\frac{Q}{2}\alpha\right) \int_{-\infty}^{\infty} du e^{\pi iu(2\xi+iQ(1-\alpha))}\times\\
&& s_b\left(i\frac{Q}{4}(1+\alpha)-\frac{\xi}{2}+u\pm\frac{m_F}{2}\right)s_b\left(i\frac{Q}{4}(1+\alpha)-\frac{\xi}{2}-u\mp \frac{m_F}{2}\right).\eean
Here $m_F$ denotes again the fugacity for the $SU(2)$ symmetry acting on the two flavors. 
The first term on the rhs represents the contribution from the singlet $\CX_1$. Our claim is now a straightforward consequence of the results presented in \cite{Collinucci:2017bwv}, where it was shown that applying twice the pentagon identity for the double sine function (see e.g. \cite{Dimofte:2011ju}) the partition function (without the contribution from $\CX_1$) is identical to that of three chiral multiplets of charge $1-\alpha$ (under (\ref{trial})) and one chiral of charge $2\alpha$. More precisely, we find the identity 
$$\mathcal{Z}=s_b\left(iQ\alpha-i\frac{Q}{2}-2\xi\right)s_b\left(2\xi+i\frac{Q}{2}-iQ\alpha\right)s_b\left(i\frac{Q}{2}\alpha-\xi\right)s_b\left( \pm m_F-\xi+i\frac{Q}{2}\alpha\right)$$ 
where we recognize in the last two terms the contribution of an $SU(2)$ adjoint with charge $1-\alpha$ under (\ref{trial}) and real mass $\xi$ under the unbroken $U(1)$ symmetry $H-C+T$. The first two terms cancel out simply because of the identity $s_b(x)s_b(-x)=1$, which is manifest from the definition of the double sine function. We thus conclude that the partition function of the monopole deformed $T(SU(2))$ is equivalent to that of an $SU(2)$ adjoint. From this observation it is clear that the partition function is extremized at $\alpha=\frac{1}{2}$, contrary to the $\CN=4$ case in which $\alpha=0$. 

We would now like to make the following observation: instead of the $SU(2)$ fugacity $m_F$ we could have used two fugacities $m_{1,2}$ satisfying the relation $m_1+m_2=0$. By formally dropping this constraint, the partition function picks a phase 
$$\mathcal{Z}\longrightarrow e^{\pi im(2\xi+iQ(1-\alpha))}\mathcal{Z},$$ 
where $m=\frac{m_1+m_2}{2}$. This fact can be simply understood as a shift of the integration variable in the partition function. This observation will be relevant below.

We now set up the inductive step. To this purpose, it is useful to notice that $T(SU(N+1))$ is equivalent to a $U(N)$ gauge theory with $N+1$ flavors and coupled to $T(SU(N))$. Using \eref{pftsun} we then conclude that the partition function of the monopole deformed $T^M(SU(N+1))$ theory can be written as follows: 
\begin{eqnarray}\label{indz}
\mathcal{Z}_{T^M(SU(N+1))}&=&\frac{s_b^{(\CX_{N})}}{N!}\int\prod_{i=1}^{N}du_ie^{2\pi i(\xi+i\frac{Q}{2}(1-\alpha))(\sum_iu_i)}\mathcal{Z}_{T^M(SU(N))}(u_i,\xi)\times \\ 
&& \frac{\prod_{i,j}s_b\left(u_i-u_j+\xi-i\frac{Q}{2}\alpha\right)\prod_{i=1}^{N}\prod_{j=1}^{N+1}s_b\left(i\frac{Q}{4}(1+\alpha)\pm u_i\mp m_j-\frac{\xi}{2}\right)}{\prod_{i<j}^{N}s_b\left(i\frac{Q}{2}\pm(u_i-u_j)\right)}\nonumber
\end{eqnarray}
where $m_j$ (subject to the constraint $\sum_jm_j=0$) denote real masses associated with the Higgs Branch $SU(N+1)$ symmetry rotating the $N+1$ flavors and $s_b^{(\CX_{N})}$ is the contribution from the singlet $\CX_{N}$, which reads 
$$s_b^{(\CX_{N})}=s_b\left(-Ni\frac{Q}{2}-(N+1)\xi+i\frac{Q}{2}(N+1)\alpha\right).$$
As explained above, once we have turned on (\ref{WdefA1}), the only unbroken $U(1)$ symmetry for which we can turn on a real mass is $C-H-\sum_iT_i$ (apart from the HB $SU(N+1)$ symmetry rotating the $N+1$ flavors), so the corresponding real mass and the $N$ FI parameters are identified. This is the reason why the parameter $\xi$ enters in $\mathcal{Z}_{T^M(SU(N))}$ as well.

By induction, we have the identity $$\mathcal{Z}_{T^M(SU(N))}=e^{(N-1)\pi i(\xi+i\frac{Q}{2}(1-\alpha))(\sum_iu_i)}\prod_{i\neq j}s_b\left(u_i-u_j-\xi+i\frac{Q}{2}\alpha\right)s_b^{N-1}\left(i\frac{Q}{2}\alpha-\xi\right)$$ where we included the phase mentioned before due to the fact that the fugacities $u_i$ do not satisfy the constraint $\sum_iu_i=0$. Plugging this in (\ref{indz}), we find that the contribution from $\mathcal{Z}_{T^M(SU(N))}$ neatly cancels against the contribution from the adjoint in the $\CN=4$ $U(N)$ vector multiplet, leaving just one singlet $\varphi$ of charge $1+\alpha$ under (\ref{trial}). This is simply because of the identity $s_b(x)s_b(-x)=1$. Therefore, the final result for $\mathcal{Z}_{T^M(SU(N+1))}$ is 
\begin{eqnarray}\label{indsqcd}
\frac{1}{N!}\int\prod_{i=1}^{N}du_i &&e^{(N+1)\pi i(\xi+i\frac{Q}{2}(1-\alpha))(\sum_iu_i)}s_b\left((N+1)\left(i\frac{Q}{2}\alpha-\xi\right)-Ni\frac{Q}{2}\right)\times \nn \\ 
&& s_b\left(\xi-i\frac{Q}{2}\alpha\right)\frac{\prod_{i=1}^{N}\prod_{j=1}^{N+1}s_b\left(i\frac{Q}{4}(1+\alpha)\pm u_i\mp m_j-\frac{\xi}{2}\right)}{\prod_{i<j}^{N}s_b\left(i\frac{Q}{2}\pm(u_i-u_j)\right)}\nonumber \\
= \frac{1}{N!}\int\prod_{i=1}^{N}du_i &&e^{(N+1)\pi i(\xi+i\frac{Q}{2}(1-\alpha))(\sum_iu_i)}s_b^{(\CX_N)} s_b^{(\varphi)}\times \\ 
&& \frac{\prod_{i=1}^{N}\prod_{j=1}^{N+1}s_b\left(i\frac{Q}{4}(1+\alpha)\pm u_i\mp m_j-\frac{\xi}{2}\right)}{\prod_{i<j}^{N}s_b\left(i\frac{Q}{2}\pm(u_i-u_j)\right)} \nn
\end{eqnarray}
where
\be
s_b^{(\varphi)} = s_b\left(\xi-i\frac{Q}{2}\alpha\right)~.
\ee

We can now observe that (\ref{indsqcd}) can be interpreted as the partition function of a $U(N)$ theory with $N+1$ flavors, two singlets ($\CX_{N}$ and $\varphi$) and superpotential $V^++\CX_{N}V^-$. Notice that this theory actually has a $SU(N)^2$ global symmetry rotating $Q$'s and $\widetilde{Q}$'s independently and we are considering real masses only for their diagonal combination, under which $Q_i$ and $\widetilde{Q}_i$ have opposite charge. 

The desired conclusion can now be obtained simply by exploiting the monopole duality (\ref{dualconfine}). At the level of $S_b^3$ partition functions, the result follows by noticing that (\ref{indsqcd}) (with the contributions from $\CX_{N}$ and $\varphi$ removed) is equivalent to the lhs of equation (8.7) of \cite{Benini:2017dud}, once we impose on the fugacities $\mu_a$ the constraint $\mu_a=\frac{\xi}{2}+i\frac{Q}{4}(1-\alpha)$ for every $a$\footnote{At first sight it might look strange to trade a real mass such as $\mu_a$ for a complex parameter. However, this is just a manifestation of the fact that we are mixing the axial $U(1)$ (in our notation $H-C$, which assigns charge $1/2$ to all $Q$'s and $\widetilde{Q}$'s) with the $R$-symmetry. As we have already explained around (\ref{compmass}), this operation is precisely equivalent to ``complexifying'' the real mass.}, we set $x_i=-u_i$ and we identify the fugacities $M_a$ with $m_j$ appearing in (\ref{indsqcd}). Using the integral identity (8.7) of \cite{Benini:2017dud} (notice that in the case $N_f=N_c+1$ we should neglect the last line of the integral identity), we then conclude that 
\be\mathcal{Z}_{T^M(SU(N+1))}=s_b^{(\CX_{N})}s_b\left(i\frac{Q}{2}N+(N+1)\left(\xi-i\frac{Q}{2}\alpha\right) \right) s_b^{(\varphi)}\prod_{i,j=1}^{N+1}s_b\left(m_i-m_j-\xi+i\frac{Q}{2}\alpha\right).\ee 
The result can be simplified by noticing that the first two terms cancel out and $s_b^{(\varphi)}$ cancels against one of the Cartan components of the meson, leaving just an adjoint of $SU(N)$ with trial $R$-charge $1-\alpha$.
This is precisely the desired conclusion: 
\begin{quote}
	The $S_b^3$ partition function of the monopole deformed $T(SU(N))$ theory is identical to that of a chiral multiplet in the adjoint of $SU(N)$. 
\end{quote}

It is also instructive to look at the mirror dual theory, in which $U(1)_C$ and $U(1)_H$ are interchanged. This amounts to flipping the sign of $\alpha$ in (\ref{trial}). As is well known, $T(SU(N))$ is self-mirror and the partition function is symmetric under exchange of FI parameters and $SU(N)$ real masses \cite{Benvenuti:2011ga}. As we have already mentioned, the monopole operators appearing in (\ref{WdefA1}) are mapped to meson components in the mirror theory. In particular, the monopoles with magnetic flux $(1,0,\dots0)$ under a single gauge group are mapped (in our convention) to the off-diagonal meson components $\widetilde{Q}_iQ^{i+1}$. This forces us to mix the $R$-symmetry with a certain combination of the Cartan components of the (now broken) $SU(N)$ Higgs branch (HB) symmetry. Specifically, the generator which replaces $\sum_iT_i$ in (\ref{trial}) is 
\be\rho=\diag\left(\frac{N-1}{2},\frac{N-3}{2}, \cdots, -\frac{N-3}{2} , -\frac{N-1}{2}\right),\ee 
and the trial $R$-symmetry becomes 
\be\label{trial2}R_{\alpha}=C+H-\alpha(C-H)+(1-\alpha)\rho.\ee 
The $S_b^3$ partition function of the deformed $T(SU(N))$ theory reads 
\begin{eqnarray} \label{massdeftn} &&\frac{\prod_{n=1}^{N-1}s_b^{(\CX_n)}}{(N-1)!}\int\prod_{j=1}^{N-1}du_je^{2\pi i \xi_{N-1}(\sum_ju_j)}\prod_{i,j}s_b\left(u_i-u_j-m+i\frac{Q}{2}\alpha\right)\times \\
&& \frac{\prod_j\prod_{k=1}^{N}s_b\left(u_j+\left(m+i\frac{Q}{2}(\alpha-1)\right)\frac{N-2k}{2}\right) s_b\left(-u_j+\left(m+i\frac{Q}{2}(\alpha-1)\right)\frac{2k-2-N}{2}\right)}{\prod_{i<j}^{N-1}s_b\left(i\frac{Q}{2}\pm(u_i-u_j)\right)}\dots\nonumber
\end{eqnarray}
In this formula $\xi_{N-1}$ denotes the FI parameter for the $U(N-1)$ gauge group, $m$ is the real mass associated with the symmetry $H-C-\rho$ and, analogously to the previous case, we are not allowed to turn on any other real masses for the HB $SU(N)$ symmetry since it is broken. The contribution from the singlet $\CX_n$, whose trial $R$-charge is $(n+1)(1-\alpha)$, reads 
$$s_b^{(\CX_n)}=s_b\left((n+1)\left(m+i\frac{Q}{2}\alpha\right)-in\frac{Q}{2}\right)$$ 
and in the second line we included the contribution of the $N$ fundamentals of $U(N-1)$. The dots stand for all other terms appearing in the partition function. We omit them since they do not play any role in our discussion. Exploiting again the identity $s_b(x)s_b(-x)=1$, we can simplify the second line which reduces to 
$$\frac{\prod_{j=1}^{N-1}s_b\left(u_j-\left(m+i\frac{Q}{2}(\alpha-1)\right)\frac{N}{2}\right) s_b\left(-u_j-\left(m+i\frac{Q}{2}(\alpha-1)\right)\frac{N}{2}\right)}{\prod_{i<j}^{N-1}s_b\left(i\frac{Q}{2}\pm(u_i-u_j)\right)}\dots$$ 
Taking this fact into account, we can notice that the partition function becomes identical to that of the linear quiver 
\be \gnode{\cap}{1} \arr{}{} \gnode{\cap}{2} \arr{}{} \cdots  \arr{}{} \gnode{\overset{\Phi}{\cap}}{N-1} \alr{q}\sqgrnode{}{1} \ee
consistently with the expectation that $N-1$ flavors at the end of the quiver became massive. The assignment of quantum numbers are compatible with the superpotential (\ref{massdef}) 
$$\W=\tilde{q}\Phi^Nq+\sum_{n=1}^{N-1}\CX_j\tilde{q}\Phi^{N-n-1}q+\ldots,$$ 
where $\Phi$ is the $U(N-1)$ adjoint and $q$, $\tilde{q}$ denote the $U(N-1)$ fundamental flavor. The matter content and interactions (denoted by $\ldots$) of the rest of the quiver is compatible with $\CN=4$ supersymmetry. 

The equality of the $S^3_b$ partition functions of the mirror theories can be understood as a consequence of the fact that $T(SU(N))$ is self-mirror: if $\xi_i$ denote the $N-1$ FI parameters of $T(SU(N))$, we can change variable and consider the $N$ parameters $e_i$ defined by the relation 
\be\label{FIrel}\xi_i=e_i-e_{i+1} ~(\text{with $i=1,\ldots, N-1$} );\qquad \sum_{i=1}^N e_i=0.\ee 
The statement that $T(SU(N))$ is self-mirror implies that 
\be\label{mirrsym}\mathcal{Z}_{T(SU(N))}(m_A,e_i,m_j)=\mathcal{Z}_{T(SU(N))}(-m_A,m_j,e_i),\ee 
where $m_A$ is the real mass for the axial symmetry $H-C$. Explicitly, the expressions on the left and right hand sides are (cf \eref{pftsun})
\bea
& \mathcal{Z}_{T(SU(N))} (m_A,e_i,m_j) \nn \\ &=\frac{1}{(N-1)!}\int\prod_{i=1}^{N-1}du_ie^{2\pi i(e_{N-1}-e_{N})(\sum_iu_i)}\mathcal{Z}_{T(SU(N-1))}(u_i,e_1,\ldots, e_{N-1})\times \nn \\ 
& \qquad \frac{\prod_{i,j}s_b\left(u_i-u_j+m_A\right)\prod_{i=1}^{N-1}\prod_{j=1}^{N}s_b\left(i\frac{Q}{4} \pm u_i\mp m_j-\frac{m_A}{2}\right)}{\prod_{i<j}^{N-1}s_b\left(i\frac{Q}{2}\pm(u_i-u_j)\right)}
\eea
and
\bea
&\mathcal{Z}_{T(SU(N))} (-m_A,e_i,m_j) \nn \\ &=\frac{1}{(N-1)!}\int\prod_{j=1}^{N-1}du_je^{2\pi i (m_1-m_2)(\sum_ju_j)}\prod_{i,j}s_b\left(u_i-u_j-m_A\right)\times \nn \\
& \qquad  \frac{\prod_j^{N-1}\prod_{i=1}^{N}s_b\left(i \frac{Q}{4} \pm u_j \mp e_i +\frac{m_A}{2}\right) }{\prod_{i<j}^{N-1}s_b\left(i\frac{Q}{2}\pm(u_i-u_j)\right)}\dots~,
\eea
where the term in the numerator in the last line denotes the $N$ flavours of fundamental hypermultiplets under the gauge group $U(N-1)$ in quiver \eref{TSUN} and the term in the denominator denotes the $U(N-1)$ vector multiplet.  The terms collected in $\cdots$ denotes the contribution from the rest of the quiver.

The symmetry under exchange of $e_i$ with $m_j$ was proven analytically for $m_A=0$ in \cite{Benvenuti:2011ga} and from the explicit expression for the partition function of $T(SU(N))$ found in the same paper, it is clear that this holds also for complex $e_i$ and $m_j$. This is expected since promoting the parameters to complex variables is interpreted as mixing of the corresponding symmetries with the $R$-symmetry.  

Exploiting the fact that (\ref{mirrsym}) is true for generic (complex) values of $m_A$ (as was proven in \cite{Bullimore:2014awa}), we can immediately derive the equality of $S^3_b$ partition functions for our deformed $T(SU(N))$ theory and its mirror since this simply follows from a specialization of (\ref{mirrsym}): on one side the monopole superpotential breaks the topological symmetries and $H-C$ to the diagonal subgroup, therefore all the $\xi_i$ parameters and $m_A$ should be identified.  This sets the real parts of all $\xi_i$ and $m_A$ to a single parameter which we shall denote by $\xi$.  Furthermore, the new interaction terms force the mixing with the $R$-symmetry according to \eref{trial}.  According to \eref{compmass}, this implies that we should add imaginary parts $i\frac{Q}{2}(1-\alpha)$ to all $\xi_i$ and $-i\frac{Q}{2}\alpha$ to $m_A$, namely 
\be \label{fixximA} \xi_i=\xi+i\frac{Q}{2}(1-\alpha);\quad m_A=\xi-i\frac{Q}{2}\alpha~. \ee 
Using this formula together with \eref{FIrel}, we immediately find 
\be e_i=\frac{N+1-2i}{2}\left(\xi+i\frac{Q}{2}(1-\alpha)\right) \ee 
and from (\ref{mirrsym}) we conclude that 
$$\prod_{n=1}^{N-1}s_b^{(\CX_n)}\mathcal{Z}_{T(SU(N))}(\xi-i\frac{Q}{2}\alpha,\frac{N+1-2i}{2}\left(\xi+i\frac{Q}{2}(1-\alpha)\right),m_j)$$ is identical to 
$$\prod_{n=1}^{N-1}s_b^{(\CX_n)}\mathcal{Z}_{T(SU(N))}(-\xi+i\frac{Q}{2}\alpha,m_j,\frac{N+1-2i}{2}\left(\xi+i\frac{Q}{2}(1-\alpha)\right)).$$ 
One can easily see that setting $\xi=-m$ the last formula is equivalent to (\ref{massdeftn}), already at the level of the integrand.

\section{$SU(2)$ gauge theory with $N$ flavours}\label{su2mirr2} 

In this section we derive the mirror dual of $SU(2)$ SQCD with zero superpotential using the monopole duality of the previous section. We also perform several consistency checks regarding the chiral ring of the two theories. 

\subsection{The $\CN=4$ mirror pairs}
We start from the following pairs of $3d$ $\CN=4$ mirror theories
\be \label{AB}
\begin{split}
(A): &\qquad \node{}{1}- \node{}{2} - \sqwnode{}{N} \\
(B): &\qquad   \underbrace{\node{\wver{}{2}}{2}-  \node{}{2} - \cdots -  \node{\wver{}{1}}{2}}_{\text{$N-2$ $U(2)$ gauge groups}}-\node{}{1}
\end{split}
\ee
The white nodes with a label $m$ represent 3d $\CN=4$ vector multiplets in the $U(m)$ group and the black lines denote the bifundamental hypermultiplets.  For the group $SU(m)$, we indicate explicitly the label $SU(m)$ under the corresponding node.

We can obtain a similar pair of theories but with $SU(2)$ gauge group instead of $U(2)$ gauge group in theory $(A)$ as follows.  We ungauge $U(1)$ inside the $U(2)$ gauge group in $(A)$.  In $(B)$, the $U(1)$ flavour symmetry is then gauged.  Therefore, we obtain
\be \label{u1su2andmirr}
\begin{split}
(A'): &\qquad \node{}{1}- \node{}{SU(2)} - \sqbnode{}{2N} \\
(B'): &\qquad   \underbrace{\node{\wver{}{2}}{2}-  \node{}{2} - \cdots -  \node{\ver{}{1}}{2}}_{\text{$N-2$ $U(2)$ gauge groups}}-\node{}{1}
\end{split}
\ee
where the blue node with a label $m$ denotes $SO(m)$ group.

\subsection{$\CN=2$ $SU(2)$ SQCD with $N$ flavours and $W=0$ and its mirror} \label{sec:su2sqcd}
The idea now is very simple: starting from theory $(A')$ in (\ref{u1su2andmirr}) we can obtain $\CN=2$ $SU(2)$ SQCD with vanishing superpotential simply by turning on the monopole deformation (\ref{WdefA1}) at the $U(1)$ node. In other words, we exploit the dual description for monopole deformed $T(SU(2))$ described before. The CB $SU(2)$ symmetry associated with the $T(SU(2))$ in theory $(A')$ is mapped to the symmetry rotating the two flavors in theory $(B')$ and, as was remarked in the previous section, the monopole deformation is equivalent to introducing in the mirror theory the field dependent mass matrix (\ref{massmat}). By activating this deformation we then land on the duality 
\be \label{mirrpairSU2}
\begin{split}
(a'): &\qquad \gnode{}{SU(2)}\, \overset{Q}{-} \,\, \sqnode{}{2N}  \quad \text{with $W_{(a')} =0$} \\
(b'): &\qquad   \sqgrnode{}{1}  \alr{q}  \underbrace{\gnode{\overset{\phi_1}{{\bigcap}}}{2} \alr{b_1}  \gnode{\overset{\phi_2}{{\bigcap}}}{2}  \alr{b_2} \cdots  \gnode{\overset{\phi_{N-3}}{{\bigcap}}}{2}  \arr{b_{N-3}}{\tilde{b}_{N-3}}}_{\text{$N-3$ $U(2)$ gauge groups}} \,\,  \gnode{{}_{s} \overset{\overset{\chi}{\bigcap}}{\cvver{}{1}} {}_{\tilde{s}}}{\underset{\underset{\phi_{N-2}}{\bigcup}}{2}} \alr{p}\,\, \gnode{\overset{\psi_{1}}{{\bigcap}}}{1} \quad \text{with $W_{(b')}$}
\end{split}
\ee
where 
\bi
\item the grey node with a label $m$ represents a 3d $\CN=2$ vector multiplets in the $U(m)$ gauge group; 
\item for the group $SU(m)$, we indicate explicitly the label $SU(m)$ under the corresponding node;
\item the notation $\cap$ denotes a chiral multiplet in the adjoint representation;  
\item the superpotential $W_{(b')}$ for the $(b')$ theory is as follows:
\be \label{supWbp}
W_{(b')} = X q \tilde{q} + q \phi_1^2 \tilde{q}   + W^{\CN=4}_{(b')} ~.
\ee
where $W^{\CN=4}_{(b')}$ contains the cubic superpotential terms coming from $\CN=4$ supersymmetry; it includes, for example, $-\tilde{b}_1 \phi_1 b_1$~.  Here we denote the flipping field $\CX_1$ in the previous section by $X$ for the sake of brevity:
\be
X = \CX_1~.
\ee
\ei
Let us now discuss in more detail how we get $(b')$ from $(B')$:
\ben
\item We start from the theory 
\be
\begin{split}
(B'): \qquad  \sqnode{}{U(2)} \alr{\mathfrak{q}} \underbrace{\gnode{\overset{\phi_1}{{\bigcap}}}{2} \alr{b_1}  \gnode{\overset{\phi_2}{{\bigcap}}}{2}  \alr{b_2} \cdots  \gnode{\overset{\phi_{N-3}}{{\bigcap}}}{2}  \arr{b_{N-3}}{\tilde{b}_{N-3}}}_{\text{$N-3$ $U(2)$ gauge groups}} \,\,  \gnode{{}_{s} \overset{\overset{\chi}{\bigcap}}{\cvver{}{1}} {}_{\tilde{s}}}{\underset{\underset{\phi_{N-2}}{\bigcup}}{2}} \alr{p}\,\, \gnode{\overset{\psi_{1}}{{\bigcap}}}{1} 
\end{split}
\ee
and turn on the superpotential corresponding to (\ref{WdefA1}):
\be \label{WBp}
W_{(B')} =  \mathfrak{q}^2 \tilde{\mathfrak{q}}_1 + X  \mathfrak{q}^1 \tilde{\mathfrak{q}}_2+ \left[  \mathfrak{q}^1  \phi_1  \tilde{\mathfrak{q}}_1+  \mathfrak{q}^2 \phi_1 \tilde{\mathfrak{q}}_2  +\tilde{W}^{\CN=4}_{(B')} \right]~,
\ee
where the square brackets contain of the usual terms coming from $\CN=4$ supersymmetry including $ \mathfrak{q}^1  \phi_1  \tilde{\mathfrak{q}}_1+  \mathfrak{q}^2 \phi_1 \tilde{\mathfrak{q}}_2$, where $\phi_i$ is the complex scalar in the $\CN=4$ vector multiplet of the $i$-th $U(2)$ gauge group from left to right, as well as the other terms collected in $\tilde{W}^{\CN=4}_{(B')}$.
\item  The F-term $\partial_{\mathfrak{q}^2} W_{(B')} = 0$ implies that
\be
\tilde{\mathfrak{q}}_1 + \phi_1 \tilde{\mathfrak{q}}_2 = 0~.
\ee
Plugging this back to \eref{WBp}, we obtain
\be
 X  \mathfrak{q}^1 \tilde{\mathfrak{q}}_2+    \mathfrak{q}^1  \phi_1^2  \tilde{\mathfrak{q}}_2+ \tilde{W}^{\CN=4}_{(B')}
\ee
We write
\be
\tilde{q} \equiv \tilde{\mathfrak{q}}_2~, \qquad q \equiv \mathfrak{q}^1~,
\ee
and hence the new effective superpotential can be written as
\be \label{supSU2wNflv}
Xq   \tilde{q}  + q \phi_1^2 \tilde{q}  + \tilde{W}^{\CN=4}_{(B')}
\ee
This is precisely the superpotential given by \eref{supWbp}. 

In the following we denote by $\phi_i$, with $i=1, \ldots, N-2$, the adjoint fields in the $U(2)$ gauge groups from left to right and by $\phi_{N-1}$ and $\phi_N$ the adjoint field in the $U(1)$ gauge group above and on the right on the $(N-2)$-th $U(2)$ gauge group.

We shall discuss further details regarding the superpotential \eref{supSU2wNflv} in Section \ref{suNmirr2} and in Appendix \ref{sec:chiralringstab}.  In the meantime, let us proceed our discussion on the chiral ring of the theories $(a')$ and $(b')$.
\een
 
\subsection{The generators of the chiral ring}
Theory $(a')$ has a global symmetry $SU(2N) \times U(1)_A$ \cite{Aharony:1997gp, Karch:1997ux, Cremonesi:2015dja}.  The two generators of the chiral ring are (1) the basic monopole operator $Y$ and (2) the mesons
\be
M_{ij} = \epsilon_{ab} Q^a_i Q^b_j~,
\ee
They transform under the global symmetry as follows:
\be
\begin{tabular}{|c|c|c|c|}
\hline
 & $U(1)_R$ & $U(1)_A$ & $SU(2N)$ \\
\hline
$Q$ & $r$ & $1$ & $[1,0,\ldots,0]$ \\
\hline
$M$ & $2r$ & $2$ & $[0,1,0,\ldots,0]$ \\
$Y$  & $2N(1-r)-2$ & $-2N$ & $[0,0,\ldots,0]$ \\
\hline
\end{tabular}
\ee
The generators of the chiral ring $M$ and $Y$ are subject to the relations
\be \label{quantumrelap}
Y M = 0~, \qquad \epsilon^{i_1 i_2 \ldots i_{2N}} M_{i_1 i_2} M_{i_3 i_4} =0~.
\ee

Now let us turn to theory $(b')$.  Let the $R$-charges of $q$ and $\tilde{q}$ be $1-2r$:
\be
R(q) = R(\tilde{q}) =1-2r~.
\ee
Since the superpotential $W_{(b')}$ has $R$-charge $2$, we have
\be
\begin{split}
&R[\phi] \equiv R[\phi_i] =2 r~ (i=1, \ldots, N-1)~, \quad R[X] = 4r~, \\ 
&R[b] \equiv R[b_i]=R[\tilde{b}_i] =R(s)=R[\tilde{s}]= R[p] = R[\tilde{p}]= 1-r  ~.
\end{split}
\ee

Therefore, the gauge invariant operator 
\be \label{longchainbpa}
q \left(\prod_{i=1}^{N-3} b_i\right) s \tilde{s} \left(\prod_{i=1}^{N-3} \tilde{b}_{N-2-i} \right) \tilde{q}
\ee
 has $R$-charge
\be
R\left[ \eref{longchainbpa} \right] = 2N(1-r)-2~,
\ee
which is indeed the $R$-charge of the monopole operator $Y$ in theory $(a')$.   We propose that
\begin{quote}
Operator \eref{longchainbpa} in theory $(b')$ is mapped to the monopole operator $Y$ in theory $(a')$ under mirror symmetry.
\end{quote}

The monopole operators in theory $(b')$ take the form
\be \label{monobp}
V_{(\vec{m}_1; \vec{m}_{2}; \ldots;\vec{m}_{N-2}; m_{N-1}; m_N)}~,
\ee
where $\vec{m}_j$ denotes the magnetic fluxes under the $j$-th $U(2)$ gauge groups:
\be
\vec{m}_j = (m_{1, j}, m_{2,j})~, \qquad m_{1, j} \geq  m_{2,j} > -\infty~,
\ee
and $m_{N-1}, m_N \in \BZ$ denote the magnetic fluxes of the two $U(1)$ gauge groups.
The $R$-charge of the monopole operator \eref{monobp} is
\be
\begin{split}
& R[ V_{(\vec{m}_1; \vec{m}_{2}; \ldots;\vec{m}_{N-2}; m_{N-1}; m_N)} ] = \\
&(1-R[q]) \sum_{i=1}^2 |m_{i,1}| + (1-R[b]) \sum_{i, j=1}^2 |m_{i,1} -m_{j,2}| \\
& +(1-R[b])\sum_{k=2}^{N-3}  \sum_{i, j=1}^2  |m_{i,k}-m_{j,k+1} | \\
& +  (1 - R[b]) \sum_{i=1}^2 (|m_{N-1}-m_{i,N-2}| +|m_N-m_{i,N-2}|) \\
& + (1 - R[\phi])\sum_{j=1}^{N-2}  |m_{1,j}-m_{2,j}| - \sum_{j=1}^{N-2}  |m_{1,j}-m_{2,j}|~.
\end{split}
\ee

It can be seen that the set of magnetic fluxes $\{(\vec{m}_1; \vec{m}_{2}; \ldots;\vec{m}_{N-2}; m_{N-1}; m_N)\}$ such that 
\be \label{condrootSO2N}
\begin{split}
& R[ V_{(\vec{m}_1; \vec{m}_{2}; \ldots;\vec{m}_{N-2}; m_{N-1}; m_N)} ] =2r~, \\ 
& \text{with $m_{1,i} \geq m_{2,i} \geq 0$ for all $i=1,\ldots, N-2$, $m_{N-1}\geq 0$ and $m_N \geq 0$}~,
\end{split} 
\ee
are in $1-1$ correspondence with the positive roots of $SO(2N)$.  As an example, for $N=4$, the set of magnetic fluxes $(\vec{m}_1, \vec{m}_{2}, m_3, m_4)$ satisfying \eref{condrootSO2N} consists of
\be
\begin{split}
&\{\{0,0\},\{0,0\},0,1\},\{\{0,0\},\{0,0\},1,0\},\{\{0,0\},\{1,0\},0,0\}~, \\
&\{\{0,0\},\{1,0\},0,1\},\{\{0,0\},\{1,0\},1,0\},\{\{0,0\},\{1,0\},1,1\}~, \\ 
& \{\{1,0\},\{0,0\},0,0\},\{\{1,0\},\{1,0\},0,0\},\{\{1,0\},\{1,0\},0,1\}~, \\ 
&\{\{1,0\},\{1,0\},1,0\},\{\{1,0\},\{1,0\},1,1\},\{\{1,0\},\{1,1\},1,1\}~;
\end{split}
\ee
these fluxes are in $1-1$ correspondence with the $12$ positive roots of $SO(8)$.  The negative roots of $SO(2N)$ are in $1-1$ correspondence with the above magnetic charges with the sign flipped.  The Cartan elements of $SO(2N)$ are then in $1-1$ correspondence with $\tr(\phi_i)$ (with $i=1, \ldots, N-2$), $\chi$ and $\psi_1$.    

In fact, theory $(b')$ does not have a global symmetry $SO(2N)$.  Although theory $(B')$ has the Coulomb branch symmetry $SO(2N)$, this symmetry enhances to $SU(2N)$ when we arrive at theory $(b')$.   The adjoint representation of $SO(2N)$ becomes the rank-two antisymmetric representation of $SU(2N)$; the latter is realised by the monopole operator with the aforementioned fluxes, together with $\tr(\phi_i)$.  We thus propose that 
\begin{quote}
such Coulomb branch operators in theory $(b')$ are mapped to the mesons $M_{ij}$ in theory $(a')$ under mirror symmetry.
\end{quote}

Finally, let us establish the correspondence between the operator $X$ in theory $(b')$ to an operator in theory $(a')$. The $R$-charge of $X$ is theory $(b')$ is $4r$, which is equal to the that of operator which is quadratic in $M$.  Since $X$ is a singlet under the manifest $SO(2N)$ global symmetry in theory $(b')$, we expect that it is mapped to another singlet of the global symmetry  in theory $(a')$. We are thus led to identify $X$ with $\Tr M^2$ (where of course $M$ is the meson in theory $(a')$). 

\subsection{Chiral ring relations}
We have seen that operator \eref{longchainbpa} gets mapped to the monopole operator $Y$ in theory $(a')$.  Since theory $(a')$ has only one gauge group $SU(2)$ and hence contains only one basic monopole operator $Y$, mirror symmetry implies that other gauge invariant operators built out of chiral fields in theory $(b')$ must either vanish or can be written in terms of \eref{longchainbpa} in the chiral ring. In this subsection, we derive such chiral ring relations from the F-terms in theory $(b')$. 

The F-term $\partial_X W_{(b')} = 0$ implies that the gauge invariant operator 
\be \label{qqtilde}
\tilde{q}_a q^a=0~,
\ee 
where $a=1,2$ is the $U(2)$ gauge index.
In addition, $\partial_{\phi_{N -1}} W_{(b')} = 0$ and $\partial_{\phi_{N }} W_{(b')} = 0$ imply that
\be \label{ptildep}
 p_a \tilde{p}^a = s_a  \tilde{s}^a = 0~.
\ee
Since $W_{(b')}$ contains the terms $b_{N-3}\phi_{N-2} \tilde{b}_{N-3}+ p \phi_{N-2} \tilde{p}+ s \phi_{N-2} \tilde{s}$, the F-terms $\partial_{\phi_{N -2}} W_{(b')} = 0$ imply that the following $2 \times 2$ matrix equations:
\be \label{FtermphiNm2}
(\tilde{b}_{N-3})^{a'}_a (b_{N-3})^b_{a'} + p_a \tilde{p}^b + s_a \tilde{s}^b=0~.
\ee
Therefore,
\be
\tr(\tilde{b}_{N-3} b_{N-3}) =0~.
\ee
Considering the F-terms $\partial_{\phi_{i}} W_{(b')} = 0$ with $i=1, \ldots, N-2$ in a similar way, we obtain
\be
\tr(\tilde{b}_i b_i) =0~, \quad i=1, \ldots, N-2~.
\ee

To obtain further chiral ring relations, let us consider the F-terms $\partial_{\phi_1} W_{(b')} = 0$:
\be \label{Ftermphi}
 q^b (\phi_1)^{d}_a \tilde{q}_d +  q^d (\phi_1)^{b}_d \tilde{q}_a - (b_1)^{a'}_a (\tilde{b}_1)^b_{a'} = 0
\ee
Contracting the indices $a$ and $b$, we obtain
\bea \label{phiqq}
\tr( \phi_1 q \tilde{q} ) =0~.
\eea
Multiplying \eref{Ftermphi} by $q^a \tilde{q}_b$ and using \eref{qqtilde}, we obtain
\be \label{bbqq}
 (b_1)^{a'}_a (\tilde{b}_1)^b_{a'} \tilde{q}_b  q^a = 0~.
\ee

On the other hand, multiplying \eref{Ftermphi} by $(b_1)^{b'}_c (\tilde{b}_1)^a_{b'}$, we obtain
\be \label{FF1}
(b_1)^{b'}_c (\tilde{b}_1)^a_{b'} \left[ q^b (\phi_1)^{d}_a \tilde{q}_d +  q^d (\phi_1)^{b}_d \tilde{q}_a \right] -  ((\tilde{b}_1 b_1 )^2)^b_c = 0
\ee
We simplify this further in two steps as follows:
\ben
\item Multiplying \eref{Ftermphi} by $(\phi_1)^c_b$, we obtain
\be
 q^b (\phi_1)^{d}_a \tilde{q}_d (\phi_1)^c_b +\underbrace{q^d ((\phi_1)^2)^{c}_d \tilde{q}_a}_{0} - (b_1)^{a'}_a (\tilde{b}_1)^b_{a'}(\phi_1)^c_b =0~,
\ee
where the second term vanishes; this follows from $\tilde{q}_a \partial_{\tilde{q}_c} W_{(b')} =0$.  Further multiplying this by $\tilde{q}_c q^e$, we have
\be
 \underbrace{q^b (\phi_1)^{d}_a \tilde{q}_d (\phi_1)^c_b \tilde{q}_c q^e}_{0} -(b_1)^{a'}_a (\tilde{b}_1)^b_{a'}(\phi_1)^c_b  \tilde{q}_c q^e =0~,
\ee
where the first term vanishes due to \eref{phiqq}.  Now we can use this relation to simplify \eref{FF1} to be
\be \label{b1b1ta}
(b_1)^{b'}_c (\tilde{b}_1)^a_{b'}  q^d (\phi_1)^{b}_d \tilde{q}_a =  ((\tilde{b}_1 b_1 )^2)^b_c~.
\ee
\item  Multiplying \eref{Ftermphi} by $(\phi_1)^c_d$, we obtain
\be
{q^b (\phi_1)^{e}_a \tilde{q}_e (\phi_1)^c_d} +  q^e (\phi_1)^{b}_e \tilde{q}_a (\phi_1)^c_d - (b_1)^{a'}_a (\tilde{b}_1)^b_{a'} (\phi_1)^c_d =0~.
\ee
Multiplying by $q^d \tilde{q}_b$ and using \eref{qqtilde} together with \eref{phiqq}, we find that the first two terms are zero and we thus obtain
\be
(b_1)^{a'}_a (\tilde{b}_1)^b_{a'} (\phi_1)^c_d q^d \tilde{q}_b = 0~.
\ee
Applying the above equation to \eref{b1b1ta}, we arrive at
\be \label{nilpotb1b1}
((\tilde{b}_1 b_1 )^2)^b_c = 0~,
\ee
\ie~ the operator $\tilde{b}_1 b_1 $ is nilpotent.  As a consequence,
\be
\tr(\tilde{b}_1 b_1) = 0~.
\ee
\een
The F-terms $\partial_{\phi_2} W_{(b')}$ implies that
\be \label{Ftermphi2}
(b_1)^{a'}_a (\tilde{b}_1)^a_{b'} =  (\tilde{b}_2)^{a'}_{a''} (b_2)^{a''}_{b'}~.
\ee
Therefore,
\be
\begin{split}
((\tilde{b}_2 b_2 )^3)^{a'}_{d'} 
&= (b_1 \tilde{b}_1)^{a'}_{b'}  (b_1 \tilde{b}_1)^{b'}_{c'} (b_1 \tilde{b}_1)^{c'}_{d'} \\
&= (b_1)^{a'}_a ((\tilde{b}_1 b_1)^2)^{a}_{b} (\tilde{b}_1)^b_{d'} \\
& =0~,
\end{split}
\ee
where the first equality follows from \eref{Ftermphi2} and the last equality follows from \eref{nilpotb1b1}.  It can be shown inductively that the operator $\tilde{b}_k b_k$ is nilpotent:
\be
(\tilde{b}_k b_k)^{k+1} = 0 \qquad \text{for all $k=1, \ldots, N-3$ and no sum over $k$}~.
\ee

On the other hand, we see that
\be
\begin{split}
(\tilde{b}_{N-3})^{a'}_{a} (b_{N-3})^b_{a'}  p_b \tilde{p}^c s_c \tilde{s}^a &\overset{\eref{FtermphiNm2}}{=} (\tilde{b}_{N-3})^{a'}_{a} (b_{N-3})^b_{a'}  p_b \tilde{p}^c \times \\
& \qquad \qquad   \left[ -(\tilde{b}_{N-3})^{b'}_{c} (b_{N-3})^a_{b'}-  p_c \tilde{p}^b \right] \\
& \overset{\eref{ptildep}}{=}  - (\tilde{b}_{N-3})^{a'}_{a} (b_{N-3})^b_{a'}  (\tilde{b}_{N-3})^{b'}_{c} (b_{N-3})^a_{b'} p_b \tilde{p}^c \\
&~ \,= - ((\tilde{b}_{N-3} b_{N-3})^2)^b_c p_b \tilde{p}^c \\
&~\,= 0~,
\end{split}
\ee
where the last equality follows from the fact that $(\tilde{b}_{N-3} b_{N-3})^b_c$ can be viewed as a nilpotent $2 \times 2$ matrix and so one can choose a basis such that  it has a canonical form $\begin{pmatrix} 0 & 1 \\ 0 & 0 \end{pmatrix}$; it follows that 
\be \label{tbNm3bNm3}
(\tilde{b}_{N-3} b_{N-3})^2=0~
\ee
with respect to this basis and thus with respect to every basis.  This relation can be generalised to 
\be
(\tilde{b}_{N-3} \tilde{b}_{N-2} \cdots \tilde{b}_\ell) (b_\ell b_2 \cdots b_{N-3})  p \tilde{p} s \tilde{s} =0~, \qquad \ell=1, 2 \ldots, N-3~.
\ee

In addition, we have
\be
\begin{split}
& (\tilde{b}_{N-3} \tilde{b}_{N-2} \cdots \tilde{b}_1 \tilde{q} q  b_1 b_2 \cdots b_{N-3})^a_b  p_a \tilde{p}^b \\
&~\,= (\tilde{b}_{N-3} \tilde{b}_{N-2} \cdots \tilde{b}_1 \tilde{q} q  b_1 b_2 \cdots b_{N-3})^a_b  \left[ (\tilde{b}_{N-3})^{a'}_a (b_{N-3})^b_{a'}- s_a \tilde{s}^b \right] \\
&\overset{\eref{tbNm3bNm3}}{=}- (\tilde{b}_{N-3} \tilde{b}_{N-2} \cdots \tilde{b}_1 \tilde{q} q  b_1 b_2 \cdots b_{N-3})^a_b  s_a \tilde{s}^b~.
\end{split}
\ee
This gives a relation involving the generator \eref{longchainbpa} of the chiral ring.

\section{$U(N)$ and $SU(N)$ SQCD with $N+k$ flavours}\label{suNmirr2}
The generalization to $U(N)$ or $SU(N)$ gauge theories is not much harder.  Let us first discuss the case of $U(N)$. 

\subsection{The mirror of $U(N)$ SQCD}
We start with the following 3d $\CN=4$ mirror theories $(A)$ and $(B)$:
\be \label{theoriesABUN} 
\begin{split}
(A): &\qquad \node{}{1}\overset{P_1}{-} \node{}{2} \overset{P_2}{-} \cdots - \node{}{N} \overset{Q}{-}\sqwnode{}{N+k} \\
(B): &\qquad   \sqwnode{}{N} \overset{\mathfrak{q}}{-}\underbrace{\node{}{N} \overset{b_1}{-}  \node{}{N}  \overset{b_2}{-} \cdots - \node{}{N} \overset{b_{k-1}}{-}  \node{\wver{\underset{s}{}}{U(1)}}{N}}_{\text{$k$ $U(N)$ gauge groups}} \overset{p_{N-1}}{-}\node{}{N-1} \cdots \node{}{3} \overset{p_2}{-}\node{}{2}\overset{p_1}{-}\node{}{1}
\end{split}
\ee

We then deform theory $(A)$ with the monopole superpotential (\ref{WdefA1}) and, as a result, all the gauge groups in the $T(SU(N))$ tail confine leaving a chiral multiplet $\Psi$ in the adjoint of $SU(N)$. We end up with the model
\be
\gnode{\overset{\varphi_N}{\bigcap}}{N} \alr{Q} \sqgrnode{}{N+k}
\ee
with superpotential
\be \label{withflipping}
(\varphi_N)^a_b  Q^b_i \tilde{Q}^i_a  + (\varphi_N)^a_b \Psi^b_a~,
\ee
Both adjoints become massive and only the trace part $\varphi$ of $\varphi_N$ survives. The F-terms with respect to $\varphi_N$ give $\Psi^b_a = -(Q^b_i \tilde{Q}^i_a)_0$ (where $()_0$ denotes the traceless component), and hence we end up with the superpotential $\varphi Q^a_i \tilde{Q}^i_a$. Introducing now by hand a singlet $S$ which flips $\varphi$ we end up with $\CN=2$ SQCD with zero superpotential.
In conclusion, we arrive at the following theory
\be \label{theoryaUN}
(a): \qquad \gnode{}{N} \alr{Q} \sqnode{}{N+k} \quad \text{with $W_{(a)}=0$}~,
\ee

Let us now consider theory $(B)$.  The superpotential \eref{WdefA1} is mapped to the field-dependent mass matrix \eref{massmat} in theory $(B)$:
\be \label{WB1}
\begin{split}
W_{(B)} 
&=   \left( \sum_{i,j=1}^{N} M^i_{j} \tilde{\mathfrak{q}}_i \mathfrak{q}^{j} \right) + \left[\left(\sum_{i=1}^N \mathfrak{q}^i \phi_1 \tilde{\mathfrak{q}}_i \right)+\tilde{W}^{\CN=4}_{(B)}\right] \\
&=   \left(\sum^{N-1}_{i=1} \tilde{\mathfrak{q}}_i \mathfrak{q}^{i+1} \right) + \left( \sum_{i=0}^{N-2} X_{i} \sum_{j=1}^{i+1} \tilde{\mathfrak{q}}_{N-i+j-1}  \mathfrak{q}^j\right) + \left[\left(\sum_{i=1}^N \mathfrak{q}^i \phi_1 \tilde{\mathfrak{q}}_i \right)+\tilde{W}^{\CN=4}_{(B)}\right]~,
\end{split}
\ee
where the square brackets contain the cubic superpotential terms that come from $\CN=4$ supersymmetry.  We isolated the term $\left(\sum_{i=1}^N \mathfrak{q}^i \phi_1 \tilde{\mathfrak{q}}_i \right)$ out explicitly and keep the rest of the terms in $\tilde{W}^{\CN=4}_{(B)}$. The latter includes, for example, $-b_1 \phi_1 \tilde{b}_1$.  In this and the following sections, we define for convenience
\be
X_{N-1-j} = \CX_j~, \quad \text{with $j=1,\ldots, N-1$}~.
\ee

The $F$-term with respect to $\mathfrak{q}^k$, for $k=1, \ldots, N-1$, gives
\be
\begin{split}
\tilde{\mathfrak{q}}_{k} + \phi_1 \tilde{\mathfrak{q}}_{k+1} + \sum_{j=k}^{N-2} X_{j} \tilde{\mathfrak{q}}_{N-j+k}=0~.	
\end{split}
\ee
Substituting the expression for $\tilde{\mathfrak{q}}_{1}$, $\tilde{\mathfrak{q}}_{2}$, $\cdots$, $\tilde{\mathfrak{q}}_{N-1}$ into \eref{WB1} recursively, we obtain 
\be
  \mathfrak{q}^1 \phi_1^N \tilde{\mathfrak{q}}_{N}   + \left[  \sum_{j=0}^{N-2} (-1)^j (j+1) X_j  \mathfrak{q}^1 \phi^j_1 \tilde{\mathfrak{q}}_{N}  + \ldots \right] +\tilde{W}^{\CN=4}_{(B)}~,
\ee
where $\ldots$ denotes the terms with higher orders in $X_j$.  However, similarly to the discussion in Appendix A of \cite{Benvenuti:2017kud}, such terms can be eliminated from the superpotential using the $F$-terms with respect to some $X_j$; this is known as the {\it chiral ring stability}.  We are thus left with
\be
  \mathfrak{q}^1 \phi_1^N \tilde{\mathfrak{q}}_{N}   + \sum_{j=0}^{N-2} c_j X_j  \mathfrak{q}^1 \phi^j_1 \tilde{\mathfrak{q}}_{N} +\tilde{W}^{\CN=4}_{(B)}~,
\ee
for some real numbers $c_j$ that depend only on $j$. Setting 
\be
\tilde{q}= \tilde{\mathfrak{q}}_{N}~, \qquad  q= \mathfrak{q}^1 ~,
\ee
and redefining $X_j$ such that $c_j$ are absorbed into their definitions, we arrive at the theory
\be \label{theorybUN}
(b): \qquad   \sqnode{}{1} \alr{q} \underbrace{\gnode{\overset{\phi_1}{\bigcap}}{N} \alr{b_1}  \gnode{\overset{\phi_2}{\bigcap}}{N}  \alr{b_2} \cdots \arr{}{} \gnode{\overset{\phi_{k-1}}{\bigcap}}{N}}_{\text{$k-1$ $U(N)$ gauge groups}} \arr{b_{k-1}}{\tilde{b}_{k-1}} \,\, \gnode{{}_{s} \overset{}{\grvver{}{1}} {}_{\tilde{s}}}{\underset{\underset{\phi_{k}}{\bigcup}}{N}}  \,\,\,\, \arr{p_{N-1}}{\tilde{b}_{N-1}} \gnode{\overset{\psi_{N-1}}{\bigcap}}{N-1} \cdots \alr{p_2}\gnode{\overset{\psi_2}{ \bigcap}}{2}\alr{p_1}\gnode{\overset{\psi_1}{ \bigcap}}{1}
~,
\ee
with superpotential
\be \label{suptheoryb}
q   \phi_1^N \tilde{q} +\left( \sum_{i=0}^{N-2} X_i  q   \phi^i_1  \tilde{q} \right)   +\tilde{W}^{\CN=4}_{(B)}~,
\ee
 Adding a flipping term $\varphi S$ in \eref{withflipping} amounts to adding to the above superpotential the term $S \tilde{s} s$, where $S$ is the flipping field in theory $(b)$.  Hence we have 
\be \label{supUNwNpkflv}
W_{(b)} = q   \phi_1^N \tilde{q} +\left( \sum_{i=0}^{N-2} X_i  q   \phi^i_1  \tilde{q} \right)   +\tilde{W}^{\CN=4}_{(B)} + S \tilde{s} s~.
\ee

Let us comment on the superpotential \eref{supUNwNpkflv}. Although this looks very similar to that discussed in Appendix A of \cite{Benvenuti:2017kud}, an important difference is the term $q \phi_1^N \tilde{q}$. One may wonder if one could apply chiral ring stability to reduce further the term $q   \phi_1^N \tilde{q}$. We explore this possibility in Appendix \ref{sec:chiralringstab} of this paper.  Let us mention briefly here some consequences of doing so and focus on the case of $N=2$ for the sake of simplicity.  First of all, we cannot drop this term totally; however, chiral ring stability allows to trade the term $q \phi_1^2 \tilde{q}$ with a new term $\eta \tilde{q} \phi q$, where $\eta = \frac{1}{2}\tr(\phi_1)$ and $\phi$ is the traceless part of $\phi_1$.  We find the following consequences: (1) $\eta$ and $\phi$ are not forced to have the same $R$-charges; and (2) there is a possibility of an emergent $U(1)$ global symmetry in the infrared which is invisible in (and incompatible with) the tree-level Lagrangian.  Since in section \ref{sec:matchZ} we manage to match the partition functions of theories $(a)$ and $(b)$ using the $R$-charges that are {\it compatible with} \eref{supUNwNpkflv}, we choose to keep the term $q   \phi_1^N \tilde{q}$ as it is in the superpotential \eref{supUNwNpkflv} and {\it not} to reduce it further using the chiral ring stability. We believe that this provides a better motivation and justification for our choice of $R$-charges in the matching of partition functions in section \ref{sec:matchZ} than what would be in the consequence (1). Moreover, in Appendix \ref{sec:chiralringstab} we will see that the emergent $U(1)$ global symmetry can be identified with a Cartan component of the axial symmetry in SQCD under mirror symmetry, which of course does not mix with the $R$-symmetry. Hence, the emergence of this $U(1)$ global symmetry does not affect the $R$-charge assignments that we use to match the partition functions in section \ref{sec:matchZ}. We conjecture that the theory $(b)$ with our choice of superpotential \eref{supUNwNpkflv} flows to the same fixed point as the theory $(b)$ with the reduced superpotential obtained using chiral ring stability (as discussed in Appendix \ref{sec:chiralringstab}).

\subsubsection{Generators and relations of the chiral ring}
In this section we will match the chiral rings of theories $(a)$ and $(b)$. 
The $F$-term $\partial_{S} W_{(b)}=0$ implies that
\be \label{sts}
s_a \tilde{s}^a = 0~.
\ee
The $F$-terms $\partial_{X_i} W_{(b)}=0$ for $i=0, 1, \ldots, N-2$ imply that the ``dressed mesons'' are zero:
\be \label{dressedmesonszero}
\tilde{q}_a (\phi^i_1)^a_b q^b = 0~, \quad \text{for $i=0, 1, \ldots, N-2$}~.
\ee
The $F$-terms $\partial_{\phi_1} W_{(b)}=0$ imply that
\be \label{Ftermphi1}
\tilde{q}_c(\phi^{N-1})^c_a q^b+(b_1)^{a'}_a (\tilde{b}_1)^b_{a'}+\sum_{i} X_i \tilde{q}_c (\phi^{i-1})^c_a q^b = 0~, 
\ee
and so, after contracting the indices $a$ and $b$, we obtain
\be
\tilde{q}_a (\phi^{N-1}_1)^a_b q^b = -(b_1)^{a'}_a (\tilde{b}_1)^a_{a'} = -\tr(b_1 \tilde{b}_1) ~.
\ee

Moreover, the quantity $(p_{N-1})_a^{a'} (\tilde{p}_{N-1})^b_{a'}$ can be viewed as an $N \times N$ nilpotent matrix; see the discussion around (3.4)-(3.6) of \cite{Gaiotto:2008ak}.  The F-terms $\partial_{\phi_k} W_{(b)}=0$ implies that
\be\label{Ftermphik}
(\tilde{b}_k)^{c'}_a  (b_k)^b_{c'}+ (p_{N-1})^{c''}_a (\tilde{p}_{N-1})^b_{c''} + s_a \tilde{s}^b =0
\ee
Contracting the indices $a$ and $b$ and using \eref{sts} together with the nilpotency of $(p_{N-1})_a^{a'} (\tilde{p}_{N-1})^b_{a'}$, we obtain
\be 
(\tilde{b}_k)^{c'}_a  (b_k)^a_{c'} = \tr(\tilde{b}_k b_k) =0~.
\ee

\subsubsection*{Nilpotency of operators}
Multiplying \eref{Ftermphi1} by $q^a (\phi_1^\ell)^d_c$ and using \eref{dressedmesonszero}, we obtain
\be \label{phiqbb}
(\phi_1^\ell)^d_c q^a (b_1)^{b'}_a (\tilde{b}_1)^b_{b'}  = 0~.
\ee
Hence, multiplying \eref{Ftermphi1} by $(b_1)^{b'}_b (\tilde{b}_1)^c_{b'}$,  we obtain
\be
\begin{split}
((b_1 \tilde{b_1})^2)^c_a &= (b_1)^{a'}_a (\tilde{b}_1)^b_{a'} (b_1)^{b'}_b (\tilde{b}_1)^c_{b'}  \\
&=-\tilde{q}_d(\phi^{N-1})^d_a q^b (b_1)^{b'}_b (\tilde{b}_1)^c_{b'} -  \sum_{i} X_i \tilde{q}_d (\phi^{i-1})^d_a q^b (b_1)^{b'}_b (\tilde{b}_1)^c_{b'} \\
&\overset{\eref{phiqbb}}{=} 0~.
\end{split}
\ee
Thus, $b_1 \tilde{b}_1$ is nilpotent.  Using the $F$-terms $\partial_{\phi_2} W_{(b)} =0$, we obtain
\be
(b_1)^{a'}_a (\tilde{b_1})^a_{b'}  = (b_2)^{a''}_{b'} (\tilde{b}_2)^{a'}_{a''}
\ee
It thus follows that
\be
\begin{split}
((\tilde{b}_2 b_2 )^3)^{a'}_{d'} 
&= (b_1 \tilde{b}_1)^{a'}_{b'}  (b_1 \tilde{b}_1)^{b'}_{c'} (b_1 \tilde{b}_1)^{c'}_{d'} \\
&= (b_1)^{a'}_a ((\tilde{b}_1 b_1)^2)^{a}_{b} (\tilde{b}_1)^b_{d'} \\
& =0~.
\end{split}
\ee
It can be shown inductively that the operator $\tilde{b}_k b_k$ is nilpotent:
\be
(\tilde{b}_k b_k)^{k+1} = 0 \qquad \text{for all $k=1, \ldots, N-3$ and no sum over $k$}~.
\ee

As a consequence, have
\be
0 = \tr( b_{k-1} \tilde{b}_{k-1}  b_{k-1} \tilde{b}_{k-1} ) = \tr[(p_{N-1} \tilde{p}_{N-1} + s \tilde{s})^2 ]~,
\ee
and, since $\tr(p_{N-1} \tilde{p}_{N-1} p_{N-1} \tilde{p}_{N-1}) = 0$, it follows that
\be
\tr(b_{k-1} \tilde{b}_{k-1}  s \tilde{s}) =0~.
\ee

\subsubsection*{$R$-charges of various fields}
Since the $R$-charge of the superpotential is $2$, we can assign the $R$-charges of $\phi_1$ and $b_i$ to be as follows:
\be
R[\phi_i] = 1+ x~, \qquad R[b_i] = R[\tilde{b}_i] = \frac{1}{2}(1-x)~.
\ee
Since the superpotential $W_{(b)}$ contains the cubic terms coming from $\tilde{W}^{\CN=4}_{(B)}$, we have
\be
\begin{split}
&R[p] = R[\tilde{p}] = \frac{1}{2}(1-x)~, \\
&R[S] = R[\tr(\phi_\ell)] =R[\tr(\psi_m)]  = 1+x~,
\end{split}
\ee
and so, from the superpotential term $S \tilde{p} p$,
\be
R[s] = R[\tilde{s}]  = \frac{1}{2}(1-x)~.
\ee

The $N+k$ diagonal components of the mesons $M^j_i = Q^j \tQ_i$ in theory $(a)$ are mapped to $\tr(\phi_i)$, $\tr(\psi_j)$ and $S$.
It also follows that
\be
R[M^j_i] = R[\tr(\phi_\ell)] = R[\tr(\psi_m)] = R[S]= 1+x~,
\ee
and so 
\be
R[Q^i] = R[\tQ_i]  = \frac{1}{2}( 1+x)~.
\ee
The $R$-charges for the minimal monopole operators $V_\pm$ of theory $(a)$ are
\be \label{RVpm}
R[V_\pm] = (N+k)\left(1-\frac{1+x}{2} \right) -(N-1) = \frac{k+2-N}{2} -x\left( \frac{k+N}{2}\right)~.
\ee

From the superpotential term $q   \phi_1^N \tilde{q}$ of $W_{(b)}$ gives
\be
2R[q]+ NR[\phi_1] = 2 \quad \Rightarrow\quad 2R[q] +N(1+x)=2\quad \Rightarrow\quad R[q] = 1- \frac{1}{2}N(1+x)~.
\ee
Hence the $R$-charges of $q b_1 b_2 \ldots b_k s$ and $\tilde{s} \tilde{b}_k \tilde{b}_{k-1} \ldots \tilde{b}_1 \tilde{q}$ are
\be
\begin{split}
&R[q b_1 b_2 \ldots b_{k-1} s] = R[\tilde{s} \tilde{b}_{k-1} \tilde{b}_{k-2} \ldots \tilde{b}_1 \tilde{q}] \\
&= \frac{k}{2}(1-x) +1 -\frac{N}{2}(1+x) \\
&= \frac{k+2-N}{2} -x\left( \frac{k+N}{2}\right) = \eref{RVpm}~.
\end{split}
\ee
 We thus propose that the minimal monopole operators $V_\pm$ of the $U(N)$ gauge group in theory $(a)$ are mapped to the following gauge invariant quantities in theory $(b)$:
\be
\begin{split}
V_+  &\quad \longleftrightarrow \quad q b_1 b_2 \ldots b_{k-1} s~, \\
V_-   &\quad \longleftrightarrow \quad \tilde{s} \tilde{b}_{k-1} \tilde{b}_{k-2} \ldots \tilde{b}_1 \tilde{q}~.
\end{split}
\ee 

The $R$-charges of the basic monopole operators $\mathfrak{M}^\pm_i$, with $i=1, \ldots, N+k$, of the $i$-th node in theory $(b)$ are
\be
\begin{split}
R[\mathfrak{M}^\pm_i] &= 2N \left( 1-  R[b_i] \right) +(N-1)(1-R[\phi_i]) -(N-1)  \\
&=2N \left[1- \frac{1}{2}(1-x) \right] +(N-1)[1-(1+ x)]  -(N-1) \\
&= 1+x~.
\end{split}
\ee
We propose that the components $M^{i+1}_i$ and $M^i_{i+1}$ of the mesons in theory $(a)$ are mapped to these basic monopole operators:
\be \label{mapmono}
\begin{split}
M^{i+1}_i  &\quad \longleftrightarrow \quad  \mathfrak{M}^+_i \\
M^i_{i+1}  &\quad \longleftrightarrow \quad  \mathfrak{M}^-_i~.
\end{split}
\ee
On the other hand, the diagonal components of the mesons in theory $(a)$ are mapped to the scalar $\phi_1, \ldots, \phi_k, \psi_1, \ldots, \psi_N$ in theory $(b)$:
\be
\begin{split}
M^{i}_i ~\text{(no sum)}  &\quad \longleftrightarrow \quad  \phi_1, \ldots, \phi_k, \psi_1, \ldots, \psi_N~.
\end{split}
\ee

\subsection{The mirror of $SU(N)$ SQCD} 

We can now easily extract a candidate mirror dual for $SU(N)$ SQCD with $N+k$ flavors and zero superpotential.  We refer to this as theory $(a')$:
\be
(a'): \qquad \gnode{}{SU(N)} \alr{Q} \sqnode{}{N+k} \quad \text{with $W_{(a')}=0$}
\ee
To get $SU(N)$ SQCD from $U(N)$ SQCD it suffices to gauge the topological symmetry of the theory, which is mapped in the mirror theory \eref{theorybUN} to the $U(1)$ symmetry rotating the multiplets $s$ and $\tilde{s}$ with opposite charge. Performing this gauging we arrive at 
\be \label{theorybSUN}
(b'): \qquad   \sqnode{}{1} \alr{q} \underbrace{\gnode{\overset{\phi_1}{\bigcap}}{N} \alr{b_1}  \gnode{\overset{\phi_2}{\bigcap}}{N}  \alr{b_2} \cdots \arr{}{} \gnode{\overset{\phi_{k-1}}{\bigcap}}{N}}_{\text{$k-1$ $U(N)$ gauge groups}} \arr{b_{k-1}}{\tilde{b}_{k-1}} \,\, \gnode{{}_{s} \overset{\overset{S}{\bigcap}}{\cvver{}{1}} {}_{\tilde{s}}}{\underset{\underset{\phi_{k}}{\bigcup}}{N}}  \,\,\,\, \arr{p_{N-1}}{\tilde{b}_{N-1}} \gnode{\overset{\psi_{N-1}}{\bigcap}}{N-1} \cdots \alr{p_2}\gnode{\overset{\psi_2}{ \bigcap}}{2}\alr{p_1}\gnode{\overset{\psi_1}{ \bigcap}}{1}
~,
\ee
with superpotential
\be \label{suptheorybS}
W_{(b')} =  q   \phi_1^N \tilde{q} +\left( \sum_{i=0}^{N-2} X_i  q   \phi^i_1  \tilde{q} \right)   +\tilde{W}^{\CN=4}_{(B)}~,
\ee 
Notice that in this case the superpotential term $Ss\tilde{s}$ is part of the $\CN=4$ gauging. Indeed this duality constitutes a generalization of the duality discussed for $SU(2)$ SQCD in the previous section. Moreover, the comment below \eref{supUNwNpkflv} also applies here.

As before, we claim that theories $(a')$ and $(b')$ are mirror dual to each other. The matching of chiral rings works as in the previous cases so we will not discuss the details. We would like to observe that the monopole operator of SQCD is mapped to the following chain of bifundamentals 
$$q b_1 b_2 \ldots b_{k-1} s\tilde{s} \tilde{b}_{k-1} \tilde{b}_{k-2} \ldots \tilde{b}_1 \tilde{q}$$

\subsection{Matching sphere partition functions} \label{sec:matchZ}

The equivalence of the $S_3^b$ partition functions essentially follows from the analysis of section \ref{tsusect}. The partition function of theory $(A)$ with the monopole deformation turned on is 
\begin{eqnarray}\label{zundef}
\mathcal{Z}_{A^M}&=&\frac{s_b(S)}{N!}\int\prod_{i=1}^{N}du_ie^{2\pi i(\xi'+i\beta\frac{Q}{2})(\sum_iu_i)}\mathcal{Z}_{T^M(SU(N))}(u_i,\xi)\times \\ 
&& \frac{\prod_{i,j}s_b\left(u_i-u_j+\xi-i\frac{Q}{2}\alpha\right)\prod_{i=1}^{N}\prod_{j=1}^{N+k}s_b\left(i\frac{Q}{4}(1+\alpha)\pm u_i\mp m_j-\frac{\xi}{2}\right)}{\prod_{i<j}^{N}s_b\left(i\frac{Q}{2}\pm(u_i-u_j)\right)}\nonumber
\end{eqnarray}
where $\xi'$ denotes the FI parameter of the $U(N)$ gauge group, $m_j$'s are the real masses for the $SU(N+k)$ global symmetry, $\xi$ is again the real mass for the $U(1)$ symmetry $H-C+\sum_ii(N-i)T_i$ discussed in section \ref{tsusect} and $s_b(S)$ is the contribution from the singlet $S$, which reads 
$$s_b(S)=s_b\left(i\frac{Q}{2}\alpha-\xi\right).$$ 
Using the result proven in section \ref{tsusect} 
$$\mathcal{Z}_{T^M(SU(N))}=e^{(N-1)\pi i(\xi+i\frac{Q}{2}(1-\alpha))(\sum_iu_i)}\prod_{i\neq j}s_b\left(u_i-u_j-\xi+i\frac{Q}{2}\alpha\right)s_b^{N-1}\left(i\frac{Q}{2}\alpha-\xi\right)$$
we find that the contributions from $T^M(SU(N))$ and $S$ cancel against the contribution from the chiral multiplet in the adjoint and the partition function \eref{zundef} becomes 
\be\label{simplza}\mathcal{Z}_{A^M}=\frac{1}{N!}\int\prod_{i=1}^{N}du_ie^{2\pi i(\xi''+i\beta'\frac{Q}{2})(\sum_iu_i)}\frac{\prod_{i=1}^{N}\prod_{j=1}^{N+k}s_b\left(i\frac{Q}{4}(1+\alpha)\pm u_i\mp m_j-\frac{\xi}{2}\right)}{\prod_{i<j}^{N}s_b\left(i\frac{Q}{2} \pm (u_i-u_j) \right)}.\ee  
We recognize here the partition function of $U(N)$ SQCD with $N+k$ flavors, where $\xi''=\xi'+\frac{N-1}{2}\xi$ is identified with the FI parameter of the theory and $\beta'=\beta+\frac{N-1}{2}(1-\alpha)$. 

The choice of exponential prefactor in the integrand of \eref{zundef} deserves some comments: a priori to identify the correct infrared $R$-symmetry one should consider the mixing with all possible $U(1)$ symmetries in the theory, compute the trial partition function and extremize it w.r.t. the mixing parameters. In all charge conjugation invariant theories, such as $\CN=4$ theories and $\CN=2$ SQCD models discussed in this paper, we know a priori that the $R$-symmetry will not mix with topological symmetries so we do not need to extremize over them. Even if we do so, we will just find that the partition function is extremized for zero mixing coefficient and we simply recover the same result we would have found discarding the mixing with the topological symmetries. On the other hand, once we have turned on the monopole superpotential \eref{WdefA1} in the $\CN=4$ theory, the invariance under charge conjugation is lost and we cannot rule out anymore the possibility that the $R$-symmetry mixes with the surviving topological symmetries. This is precisely the reason why we introduced the parameter $\beta$ in \eref{zundef}\footnote{We would like to thank Francesco Benini for suggesting this procedure.}: in theory $(A)$ the monopole superpotential \eref{WdefA1} leaves the $U(N)$ topological symmetry $T_N$ unbroken, but since charge conjugation invariance is lost, we should consider the trial $R$-symmetry 
$$R_{\alpha,\beta}=R_{\alpha}+\beta T_N~,$$ 
with $R_\alpha$ given by \eref{trial}, and then extremize over $\beta$. Based on these considerations, \eref{WdefA1} is interpreted as the trial partition function which should be extremized. The extremization over $\beta$ can be circumvented with the following simple observation: \eref{zundef} is equivalent to \eref{simplza}, which in turn can be identified with the trial partition function of $U(N)$ SQCD with $N+k$ flavors and trial $R$-symmetry 
$$R_{\alpha,\beta}=R_{\alpha}+\beta' T,$$ 
where $T$ is the topological symmetry of the theory. Since in this theory charge conjugation is a symmetry, we know that the partition function is extremized at $\beta'=0$, or equivalently 
\be \label{valuebeta}
\beta=-\frac{N-1}{2}(1-\alpha)~.
\ee 
This is manifest in the special case $\xi=\xi'=m_j=0$, since \eref{simplza} is an even function of $\beta'$. We thus conclude that the partition function extremized over $\beta$ is identical to that of $\CN=2$ $U(N)$ SQCD with $N+k$ flavors, as we expected from our duality arguments. Notice that here charge conjugation is an accidental symmetry emerging in the IR, like the axial $SU(N+k)$ symmetry which is not present in the parent $\CN=4$ theory.

We would like to remark another important consequence of the nonzero value of $\beta$: in the original $\CN=4$ theory the monopole $V^+$ (with unit magnetic flux under $U(N)$ only) has trial $R$-charge (in the convention of section \ref{tsusect}) $\frac{k+1}{2}(1+\alpha)$. Once the monopole deformation is activated and we introduce the mixing of the $R$-symmetry with $T_{N}$, the $R$-charge of the monopole is shifted by $-\frac{N-1}{2}(1-\alpha)$ and the resulting $R$-charge is precisely that of a monopole operator in $\CN=2$ $U(N)$ SQCD with $N+k$ flavors of charge $\frac{1-\alpha}{2}$. After confinement of the gauge nodes in the tail, the monopole $V^+$ is identified with the monopole operator in $U(N)$ SQCD and the $R$-charge assignment is automatically consistent with this interpretation. 

Let us now match \eref{zundef} (or \eref{simplza}) with the partition function of theory $(b)$. The equality of the partition functions of theories $(A)$ and $(B)$ (before the monopole deformation) is a consequence of $\CN=4$ mirror symmetry: indeed theories $(A)$ and $(B)$ admit a Hanany-Witten brane realization in Type IIB and they are related by the action of S-duality on the brane system, as is expected for mirror dual theories. The matching of partition functions for many mirror theories in this class was checked analytically in \cite{Benvenuti:2011ga,Kapustin:2010xq}. 

In the case at hand the most convenient way to proceed is to notice that theory $(A)$ can be obtained via higgsing starting from $T(SU(N+k))$: it actually corresponds to $T^{\vec{\Lambda}}(SU(N+k))$ where $\vec{\Lambda}\equiv(k,1\dots1)$ denotes the partition labelling the corresponding nilpotent orbit. If we denote with $\xi_i$ the FI parameters associated with the ``balanced'' gauge groups $U(1),\dots,U(N-1)$ and with $\xi'$ the $U(N)$ FI parameter, we can introduce as in section \ref{tsusect} the $N+1$ parameters $e_i$ defined as follows: 
\be \label{xiandxip} \xi_i=e_i-e_{i+1}~ (\text{with $i=1,\ldots, N-1$});\qquad \xi'=e_N-e_{N+1}. \ee
Similarly to \eref{FIrel}, the $e_i$'s satisfy one relation which in the present case reads 
\be\label{FInews}\sum_{i=1}^{N}e_i+ke_{N+1}=0.\ee 
This constraint was derived in \cite[(3.14)]{Cremonesi:2014kwa} in the context of the Hilbert series\footnote{The parameters $x_i$ appearing in (3.14) of \cite{Cremonesi:2014kwa} are fugacities and actually correspond to the exponentials of the parameters $e_i$ used in the present work. This is the reason why the constraint among $e_i$'s involves sums instead of products.} and generalizes \eref{FIrel} which holds in the case of trivial nilpotent orbit $\vec{\Lambda}=(1,\dots,1)$. As was pointed out in the same reference, the parameters $e_i$ describe the contribution from the various NS5 branes and so should be identified with real masses of the various flavors in the mirror theory, in which NS5 branes are replaced by $D5$ branes. This constraint can be interpreted as saying that the ``real masses'' associated with the cartan generators of the $SU(N)$ topological symmetry are $\tilde{e}_i\equiv e_i+\frac{k}{N}e_{N+1}$ for $1\leq i\leq N$. These indeed satisfy the relation $\sum_i\tilde{e}_i=0$. 

At the level of partition functions, the statement of mirror symmetry is
\be\label{partgen} \mathcal{Z}_A(m_A,e_i,m_j)=\mathcal{Z}_B(-m_A,m_j,e_i),\ee 
where $m_A$ is the real mass for the ``axial'' symmetry $H-C$, the parameters $e_i$ are interpreted as (linear combinations of) FI parameters in theory $(A)$ and as real masses for the $SU(N)\times U(1)$ symmetry in theory $(B)$. $m_j$ denote of course real masses for the $SU(N+k)$ symmetry in theory $(A)$ and FI parameters in theory $(B)$. Note that \eref{partgen} was proven in \cite{Bullimore:2014awa} for a general value of $m_A$\footnote{We thank Sara Pasquetti for pointing this out to us. In the case $m_A=0$ \eref{partgen} follows from the results of \cite{Benvenuti:2011ga,Kapustin:2010xq}. For $T(SU(2))$, this statement is also proven in \cite{Collinucci:2017bwv}.}.
Equation \eref{partgen} implies the equality between the following two parition functions (as in \eref{partgen}):
\bea
\mathcal{Z}_{A} (m_A,e_i,m_j) &=\frac{1}{N!}\int\prod_{i=1}^{N}du_ie^{2\pi i(e_{N}-e_{N+1})(\sum_iu_i)}\mathcal{Z}_{T(SU(N))}(u_i,e_1,\ldots, e_N)\times \nn \\ 
& \frac{\prod_{i,j}s_b\left(u_i-u_j+m_A\right)\prod_{i=1}^{N}\prod_{j=1}^{N+k}s_b\left(i\frac{Q}{4} \pm u_i\mp m_j-\frac{m_A}{2}\right)}{\prod_{i<j}^{N}s_b\left(i\frac{Q}{2}\pm(u_i-u_j)\right)}
\eea
and
\bea \label{ZBexp}
\mathcal{Z}_{B} (-m_A,e_i,m_j) &=\frac{1}{N!}\int\prod_{j=1}^{N}du_je^{2\pi i (m_1-m_2)(\sum_ju_j)}\prod_{i,j}s_b\left(u_i-u_j-m_A\right)\times \nn \\
& \frac{\prod_j\prod_{i=1}^{N}s_b\left(i \frac{Q}{4} \pm u_j \mp e_i +\frac{m_A}{2}\right) }{\prod_{i<j}^{N}s_b\left(i\frac{Q}{2}\pm(u_i-u_j)\right)}\dots~,
\eea
where the terms in the numerator in the last line correspond to the hypermultiplet $\mathfrak{q}$ of theory $(B)$ in \eref{theoriesABUN} and the term in the denominator corresponds to the contribution of the leftmost $U(N)$ vector multiplet.  The term $\cdots$ denotes the contribution from the rest of quiver $(B)$ in \eref{theoriesABUN}.

The desired result simply follows from a specialization of \eref{partgen} by setting (for $1 \leq j\leq N$)
\bea \label{mAej}
m_A &=\xi-i\frac{Q}{2}\alpha~;\\ 
e_j&=\frac{k}{N+k}\left(\xi'+i\frac{Q}{2}\beta\right)+\left(N-j-\frac{N^2-N}{2N+2k}\right)\left(\xi+i\frac{Q}{2}(1-\alpha)\right),
\eea
and 
\be \label{eNp1}
e_{N+1}=-\frac{N}{N+k}\left(\xi'+i\frac{Q}{2}\beta\right)-\frac{N^2-N}{2N+2k}\left(\xi+i\frac{Q}{2}(1-\alpha)\right)~.
\ee 
These formulas can be obtained in a similar way to the discussion around \eref{fixximA}: solving simultaneously the system of equations \eref{xiandxip} and \eref{FInews} and identifying all the FI parameters $\xi_i$ with the real mass $m_A$ for the ``axial'' symmetry to a single parameter $\xi$, we obtain the real parts of $m_A$, $e_j$ (for $j=1,\ldots, N$), and $e_{N+1}$ as above.  The imaginary parts are fixed by the consistency with the aforementioned trial $R$-symmetry
\be R_{\alpha,\beta}=R_{\alpha}+\beta T_N ,\ee
with $R_\alpha$ given by \eref{trial}.  According to \eref{compmass}, this implies that we should add imaginary parts $i\frac{Q}{2}(1-\alpha)$ to all $\xi_i$ (with $i=1,\ldots, N-1$), $i \frac{Q}{2} \beta$ to $\xi'$, and $-i\frac{Q}{2}\alpha$ to $m_A$.  Solving again \eref{xiandxip} and \eref{FInews}, we obtain the imaginary parts of the above results.

Once we introduce the contribution of the singlets $S$ and $X_i$ and extremize w.r.t. $\beta$, the left hand side of \eref{partgen} reduces to the partition function of $U(N)$ SQCD in the way that have already discussed around \eref{simplza}.  The right hand side instead, in which $e_i$'s represent real masses for the flavors, reduces to $(b)$: because of the choice made above for the parameters $e_i$, the contributions from $N-1$ out of the $N$ flavors at the end of the quiver cancel out thanks to the identity $s_b(x)s_b(-x)=1$.  In particular, it can be seen from \eref{mAej} that the terms $s_b\left(i \frac{Q}{4} +u_j -e_i+ \frac{m_A}{2} \right)$ and $s_b\left(i \frac{Q}{4} -u_j +e_{i+1}+ \frac{m_A}{2} \right)$, with $i=1,\ldots, N-1$, in \eref{ZBexp} cancel each other. In conclusion, we are left with one flavor $q$ and $\tilde{q}$, corresponding to the terms $s_b\left(i \frac{Q}{4} +u_j -e_{N}+ \frac{m_A}{2} \right)$ and $s_b\left(i \frac{Q}{4}-u_j +e_{1}+ \frac{m_A}{2} \right)$, whose trial $R$-charge is given by 
\bea
R_{\alpha,\beta}(q) &=1+\frac{k\beta}{N+k}-(1-\alpha)\frac{N^2+k}{2N+2k}~,\\ R_{\alpha,\beta}(\tilde{q}) &=1-\frac{k\beta}{N+k}-(1-\alpha)\frac{N^2+2Nk-k}{2N+2k}~,
\eea
where these values can be easily extracted from the term $i \frac{Q}{2} (1-R_{\alpha,\beta}(q,\tilde{q}))$ inside the argument of $s_b$.

The singlets $\tilde{s}$ and $s$ appearing in $(b)$ correspond to the terms $s_b\left(i \frac{Q}{4} -u_j +e_{N+1}+ \frac{m_A}{2} \right)$ and $s_b\left(i \frac{Q}{4}+u_j -e_{N+1}+ \frac{m_A}{2} \right)$ in \eref{ZBexp}.
Their $R$-charges are 
\be
R_{\alpha,\beta}(s,\tilde{s})=\frac{1+\alpha}{2}\mp\left((1-\alpha)\frac{N(N-1)}{2N+2k}+\frac{N\beta}{N+k}\right),
\ee 
with $-$ for $s$ and $+$ for $\tilde{s}$. This assignment of $R$-charge is compatible with all the superpotential terms appearing in $(b)$, but as we can notice it is not compatible with charge conjugation invariance since chiral multiplets with and without tilde have different trial $R$-charge. This fits perfectly with the previous discussion for SQCD: the off-diagonal mass terms forces the mixing with a baryonic symmetry and breaks charge conjugation invariance, so we should extremize over all surviving baryonic symmetries as well. However, charge conjugation reappears as an accidental symmetry in the IR and this immediately tells us that the trial partition function will be extremized for the value of $\beta$ which sets to zero the difference between the $R$-charge of fields with and without tilde. Imposing $R_{\alpha,\beta}(q)=R_{\alpha,\beta}(\tilde{q})$ and $R_{\alpha,\beta}(s)=R_{\alpha,\beta}(\tilde{s})$ we find
\be \beta=-\frac{N-1}{2}(1-\alpha)~,\ee 
in perfect agreement with \eref{valuebeta} of the mirror side. 

\subsubsection*{The case of $SU(N)$ gauge theory}
The above discussion can be easily generalized to the case of $SU(N)$ SQCD with $N+k$ flavors: it is enough to gauge the topological symmetry $T_N$ (or its baryonic counterpart in the mirror theory). This has the effect of removing the central $U(1)$ inside $U(N)$ in theory $(A)$ and gauge the $U(1)$ symmetry acting on $s$ and $\tilde{s}$ fields in theory $(B)$ (notice that this gauging combined with the superpotential term $Ss\tilde{s}$ produces an $\CN=4$ gauging). At the level of partition functions, this just amounts to integrating over the parameter $\xi'$ in \eref{partgen}. We have as before 
$$m_A=\xi-i\frac{Q}{2}\alpha$$ but we have only $N$ parameters $e_i$, with $i=1,\ldots,N$, satisfying the constraint 
$$\sum_{i=1}^N e_i=0$$ as in section \ref{tsusect}. We thus have as in the $T(SU(N))$ case 
$$e_i=\frac{N+1-2i}{2}\left(\xi+i\frac{Q}{2}(1-\alpha)\right).$$ 
The parameter $\beta$ does not arise this time: technically this is due to the fact that (in theory $(A)$) the integration over $\xi'$ sets to zero the sum of the integration variable, reproducing the correct Haar measure for $SU(N)$. This directly removes the phase coming from the $T(SU(N))$ tail. This result is indeed expected physically, because in a $SU(N)$ gauge theory there is no topological symmetry which can possibly mix with the $R$-symmetry.

\subsubsection*{A comment on the flavour symmetry}
A $U(N)$ gauge theory, {\it resp.} $SU(N)$ gauge theory, with $N+k$ flavours and zero superpotential has a flavour symmetry $SU(N+k) \times SU(N+k) \times U(1)_A$, resp. $SU(N+k) \times SU(N+k) \times U(1)_A \times U(1)_B$. However, from the perspective of the mirror theory $(b)$ in \eref{theorybUN}, {\it resp.} theory $(b')$ in \eref{theorybSUN}, we see that the number of $U(1)$ topological symmetries is $N+k-1$, {\it resp.} $N+k$.  Thus, not all Cartan elements of $SU(N+k) \times SU(N+k)$ are visible in the mirror theory; only those of the diagonal subgroup are manifest in the quiver description. In other words, the $SU(N+k) \times SU(N+k)$ symmetry is ``hidden'' in the mirror theory and only arises at low energies.  In Appendix \ref{sec:chiralringstab}, we discuss about the possibility that there may exist an extra $U(1)$ global symmetry that emerges in the infrared due to the chiral ring stability condition.  Nevertheless, this does not explain the remaining hidden Cartan elements.  It would be interesting to get further insight on this point in future work.

The symmetry enhancement can also be seen from the perspective of the partition function: the $\CN=4$ theory has the $SU(N+k)$ flavour symmetry but, as remarked below \eref{indsqcd}, once the singlet is flipped this symmetry enhances to $SU(N+k) \times SU(N+k)$.   In the partition function of the $\CN=4$ theory, one can turn on real masses for only the diagonal combination of $SU(N+k) \times SU(N+k)$ and these are mapped to
FI parameters in the mirror theory.  Indeed, when the adjoint field is removed, one is allowed to introduce real masses for both $SU(N+k)$ symmetries and these should correspond to a ``doubling'' of the FI parameters on the dual side.  It would be nice to get a better understanding of this ``doubling'' in the mirror theory in the future.

\section{Brane realisation} \label{sec:branes}
\begin{figure}[htbp]
\begin{center}
\includegraphics[scale=0.4]{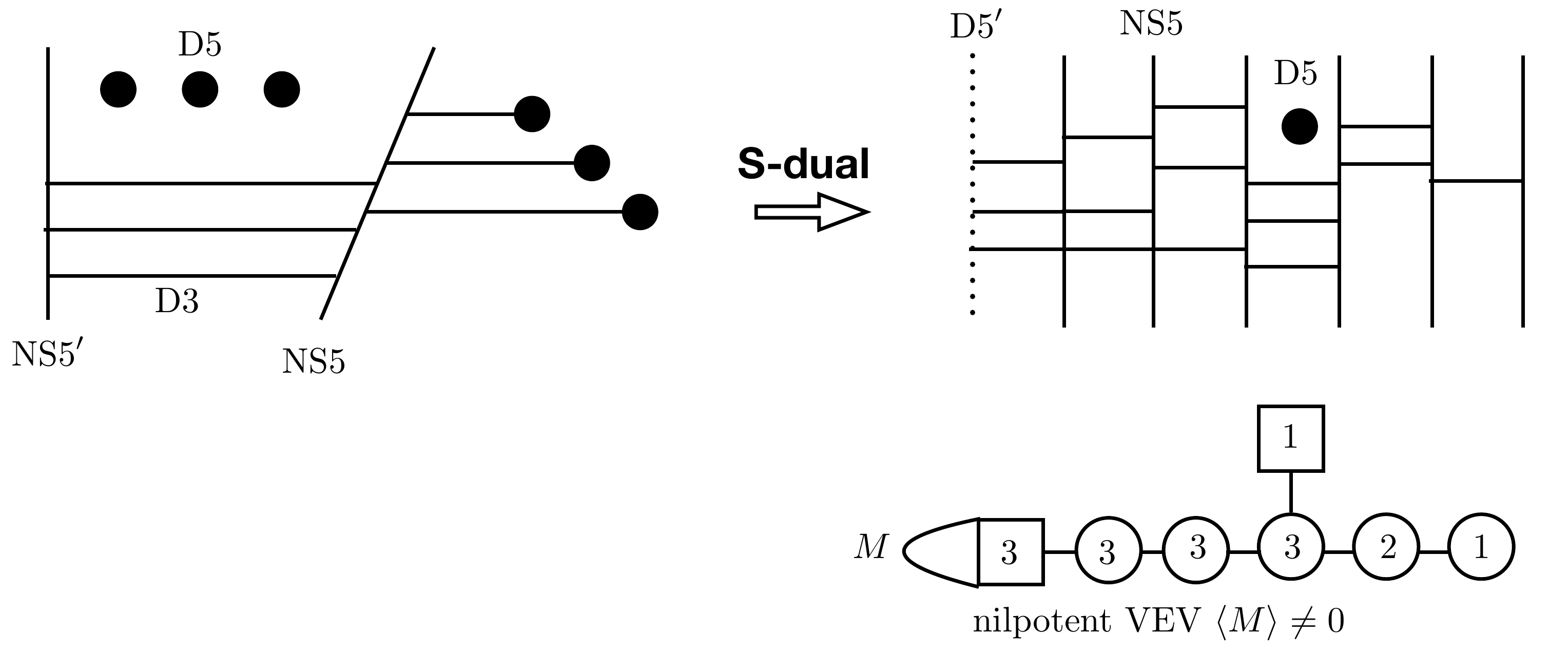}
\caption{The brane realisation of mirror pairs \eref{theoryaUN} and \eref{theorybUN} with $N=3$ and $k=3$.}
\label{brane1}
\end{center}
\end{figure}

In this section, we discuss a brane realisation \cite{Hanany:1996ie, Elitzur:1997fh, Elitzur:1997hc, Giveon:1998sr} of the mirror pairs $(a)$ and $(b)$ given by \eref{theoryaUN} and \eref{theorybUN}.    It is instructive to describe this using a particular example, say for $N=3$ and $k=3$, depicted in Figure \ref{brane1}.  This can be generalised for any value of $N$ and $k$.

The branes in the left diagram span the following directions
\be
\begin{tabular}{c|cccccccccc}
& 0 & 1 & 2 & 3& 4 &5 &6 &7 &8 &9 \\
\hline
D3                        & X & X & X &   &     &    &X &  &     &    \\
NS5                      & X & X & X & X & X &X &   &   &    &    \\
$\mathrm{NS5}'$ & X & X & X & X &     &   &   &  & X & X \\
D5                        & X & X & X &    &     &   &   &X& X & X \\          
\end{tabular}
\ee
As described in \cite{Elitzur:1997fh, Elitzur:1997hc, Giveon:1998sr}, this corresponds to $U(N)$ SQCD with $N+k$ flavours and zero superpotential; this is indeed theory $(a)$ described in \eref{theoryaUN}.

In order to determine the mirror theory, we apply the S-duality to the brane system described above \cite{Hanany:1996ie}.   The NS5-brane becomes a D5-brane, the $\mathrm{NS5}'$-brane becomes a $\mathrm{D5}'$-brane, and  the $\mathrm{D5}$-brane becomes an $\mathrm{NS5}$-brane.  Arranging the leftmost NS5-brane in the right diagram to cut the D3-branes, we see that the motion of the leftmost D3-branes segment along the $8$ and $9$ directions corresponds to turning on the nilpotent VEV $\langle M \rangle \neq 0$.   This VEV higgses the flavour symmetry to $U(1)$ and leads to the first two terms in the superpotential \eref{suptheoryb}.  Observe that the remaining part of the brane configuration is still $\CN=4$ supersymmetric.  We thus expect the presence of the term $\tilde{W}^{\CN=4}_{(B)}$ in \eref{suptheoryb}.

This idea can be generalised to other classical gauge groups.  From the perspective of branes, this corresponds to introducing an appropriate orientifold plane to the system. We shall present such results in the following section and in Appendix \ref{sec:orthosymp}.

\section{$USp(2k)$ with $N_f$ fundamental flavours and one antisymmetric traceless chiral multiplet}\label{antiusp}
As proposed in \cite[Fig. 61, p. 139]{Gaiotto:2008ak}, we have the following 3d $\CN=4$ mirror pairs:
\be
\begin{split}
&(A): \qquad \node{}{1}-\node{}{2}-\cdots-\node{}{2k-1} - \rnode{}{2k}-\sqbnode{}{2N_f} \\
&(B): \qquad \sqwnode{}{2k}- \underbrace{\node{}{2k}-\cdots - \node{}{2k} }_{N_f-3}- \node{\ver{}{k}}{2k} -\node{}{k} 
\end{split}
\ee

We can apply a similar procedure as in the previous section and obtain the following 3d $\CN=2$ mirror pairs:

\be
\begin{split}
&(a): \qquad \rnode{\overset{A'}{\cap}}{2k} \overset{Q}{-} \sqgrnode{}{2N_f} \qquad \text{with $W_{(a)} =0$}\\
&(b): \qquad \sqgrnode{}{1} \alr{q} \gnode{\overset{\phi_1}{\cap}}{2k}  \alr{b_1} \gnode{\overset{\phi_2}{\cap}}{2k}  \alr{b_2} \cdots   \gnode{\overset{\phi_{N_f-3}}{\cap}}{2k} \alr{b_{N_f-3}} \,\,\gnode{{}_{s} \overset{\overset{\chi}{\cap}}{\grvcver{}{k}} {}_{\tilde{s}}}{\underset{\underset{\phi_{N_f-2}}{\bigcup}}{2k}} \alr{p} \gnode{\overset{\psi_1}{\cap}}{k}  \qquad \text{with $W_{(b)}$}
\end{split}
\ee
where the red node denotes the gauge group $USp(2k)$, $A'$ denotes the rank-two traceless anti-symmetric chiral multiplet, and
\be
W_{(b)} = \tilde{q} \phi^{2k} q +  \sum_{j=0}^{2k-2}  X_j \tilde{q} \phi^{j} q + W^{\CN=4}_{(b)}~.
\ee
Note that for $k=1$, we recover the mirror pair \eref{mirrpairSU2}

\subsubsection*{$R$-charges and operator maps}
Let the $R$-charges of $q$ and $\tilde{q}$ be $1-kr$:
\be
R(q) = R(\tilde{q}) =1-2kr~.
\ee
Since the superpotential $W_{(b)}$ has $R$-charge $2$, we have
\be
\begin{split}
&R[\phi] \equiv R[\phi_i] =2 r~ (i=1, \ldots, N_f)~, \\ 
&R[b] \equiv R[b_i]=R[\tilde{b}_i] =R(s)=R[\tilde{s}]= R[p] = R[\tilde{p}]= 1-r  ~.
\end{split}
\ee
Therefore, the gauge invariant operator 
\be \label{longchainbpx}
q \left(\prod_{i=1}^{N_f-3} b_i\right) s \tilde{s} \left(\prod_{i=1}^{N-3} \tilde{b}_{N_f-2-i} \right) \tilde{q}
\ee
 has $R$-charge
\be
R\left[ \eref{longchainbpx} \right] = 2N_f(1-r)+ 2 (k-1)(1- 2r) -2k~.
\ee

If we assign the $R$-charges of the fields $Q$ and $A'$ in theory $(a)$ to be
\be\label{rantis}
R[Q]= r~, \qquad R[A'] = 2r~,
\ee
then the $R$-charge of the minimal monopole operator $Y$ of the $USp(2k)$ gauge group in theory $(a)$ is (see \eg,~ \cite[Sec. 5]{Amariti:2015vwa})
\be
R[Y] = 2N_f(1-r)+ 2 (k-1)(1- 2r) -2k~.
\ee
Indeed, we propose the following operator map
\be
Y ~\longleftrightarrow ~q \left(\prod_{i=1}^{N_f-3} b_i\right) s \tilde{s} \left(\prod_{i=1}^{N-3} \tilde{b}_{N_f-2-i} \right) \tilde{q}~,
\ee
which is to be expected from mirror symmetry.  The mesons $M^{ij}=J^{ab} Q^{i}_a Q^{j}_b$, with $i,j =1, \ldots, 2N_f$, in theory $(a)$ has $R$-charge:
\be
R[M] = 2r ~,
\ee
The operator maps of each component of $M$ to the operators of theory $(b)$ are similar to those stated around \eref{monobp}--\eref{condrootSO2N}.  In particular, if we view $M$ as a matrix transforming in the adjoint representation of $SO(2N_f)$, then the Cartan elements are mapped to $\tr(\phi_i)$ (with $i=1, \ldots, N_f-2$), $\tr(\chi)$ and $\tr(\psi_1)$; and the element of the root are mapped to the minimal monopole operators in theory $(b)$, whose $R$-charge are $2r$.

Notice that we can match the chiral rings of the two theories only if we assume the R-charge assignment \eref{rantis}. Such a relation between the $R$-charges of traceless anti-symmetric and fundamental fields is not expected in general and we interpret this fact as evidence that the mirror theory $(b)$ has an emergent $U(1)$ symmetry which mixes with the $R$-symmetry. Equation \eref{rantis} is not very surprising after all: both the $USp(2N)$ adjoint and the anti-symmetric chirals originate as components of the $SU(2N)$ adjoint which emerges upon confinement of the gauge groups in the $T(SU(2N))$ tail. Because of the $\CN=4$ superpotential terms of theory $(A)$, this field is constrained to have the same $R$-charge as the meson built out of the $USp(2N)$ fundamentals, reproducing \eref{rantis}. This relation is maintained until the very last confinement step, in which a symmetry acting on the anti-symmetric chiral only emerges. In the dual theory this is just a hidden symmetry.

\section{Concluding remarks}

In this paper we have seen that in three dimensions there is a precise method to introduce a chiral multiplet in the adjoint representation of a unitary gauge group: it is enough to couple the theory to a $T(SU(N))$ theory and turn on a monopole superpotential deformation. This procedure allows to modify in a controlled way the matter content of a three-dimensional gauge theory and, as we have explained extensively, this can be used to generate new dual descriptions of $\CN=2$ SQCD. We tested our duality proposal with a variety of methods, including analysis of the chiral rings and of sphere partition functions. 

In principle our construction can be iterated coupling several $T(SU(N))$ tails and activating the monopole superpotential deformation for all of them. This has the effect of introducing several adjoint chirals. As we have illustrated in section \ref{antiusp}, the price we have to pay, if we want to use this method to introduce new matter fields rather than removing them, is the presence of accidental symmetries. One then needs to understand how to detect them. 

There are many directions worth investigating. First of all it would be interesting to obtain the analogous result for $T(SO(2N))$ theories. This would shed more light on the dualities we conjecture for orthogonal or symplectic SQCD in Appendix \ref{sec:orthosymp}. It would also be interesting to generalize our construction to the case of $\CN=2$ quiver theories, as well as to case in which tensor matter is included. Yet another interesting question is to study the reduction of the mirror pairs in this paper to two dimensions along the line of \cite{Aganagic:2001uw, Aharony:2017adm}.  This could potentially lead to new mirror theories in two dimensions that have not been studied before.

\acknowledgments{We are grateful to Antonio Amariti, Francesco Benini, Sergio Benvenuti, Bruno Le Floch, Sara Pasquetti and Roberto Valandro for valuable discussions and comments. The research of S.G. is partly supported by the INFN Research Project ST\&FI. The research of N.M. is supported by the INFN. Both authors also thank the Pollica Summer Workshop 2017, where significant progress of this research project has been made.  The workshop has been partly supported by the ERC STG grant 306260.}

\appendix

\section{Chiral ring stability and emergent global symmetries} \label{sec:chiralringstab}

In this section we discuss in more detail emergent symmetries in our mirror theories, especially in connection with the chiral ring stability criterion of \cite{Benvenuti:2017lle} (see also \cite{Benvenuti:2017kud}). 
Before proceeding with the analysis, let us briefly review the findings of \cite{Aharony:1997bx} in the abelian case. The mirror of $\CN=4$ SQED with $N$ flavors (plus a free hypermultiplet) is a circular quiver with $N$ $U(1)$ gauge groups and bifundamental hypermultiplets $q_i,\tilde{q}_i$ ($i=1\dots N$) charged under consecutive $U(1)$ groups. We denote the singlets in the vector multiplets as $\phi_i$ ($i=1\dots N$). The superpotential of the mirror theory is 
$$\mathcal{W}=\sum_{i=1}^{N}(\phi_i-\phi_{i+1})\tilde{q}_iq^i\quad (\phi_{N+1}\equiv\phi_1).$$ 
To recover $\mathcal{N}=2$ SQED we introduce a chiral multiplet $\lambda$ and couple it to the singlet in the $\mathcal{N}=4$ vector multiplet to make it massive. In the dual theory this is implemented by coupling the extra singlet (which we call again $\lambda$) to all the mesons. The superpotential becomes 
$$\mathcal{W}=\sum_{i=1}^{N}(\phi_i-\phi_{i+1})\tilde{q}_iq^i+\lambda \left(\sum_i\tilde{q}_iq^i \right).$$ 
If we now perform the following field redefinition on the singlets: 
$$S_i=\phi_i-\phi_{i+1}+\lambda\quad(i=1\dots N);\quad \phi=\sum_i\phi_i,$$ 
we find that $\phi$ drops out of the superpotential and decouples (together with the diagonal combination of the $U(1)$ vector multiplets) and we conclude that $\mathcal{N}=2$ SQED is dual to the $U(1)^N/U(1)$ theory with superpotential 
$$\mathcal{W}=\sum_iS_i\tilde{q}_iq^i.$$ 
The above field redefinition is unitary, hence the K\"{a}hler potential is not affected. This model has $N-1$ $U(1)$ topological symmetries, a baryonic symmetry and $N$ $U(1)$ symmetries under which the bifundamentals have charge 1 and the singlets $S_i$ have charge $-2$. This precisely reproduces the rank $2N$ of the global symmetry $SU(N) \times SU(N) \times U(1)_A \times U(1)_J$ of SQED with $N$ flavors.

Let's now turn to the analysis of nonabelian theories and for definiteness we focus on the simplest nontrivial case: the mirror dual of $SU(2)$ SQCD with three flavors. The arguments can easily be extended to higher rank cases. As we have argued in section \ref{sec:su2sqcd}, the mirror theory is the quiver $(b')$ in (\ref{mirrpairSU2}) with superpotential (we use the same notation) 
\be\label{sup23}\mathcal{W}=X\tilde{q}q+\tilde{q}\phi_1^2q+\chi\tilde{s}s+\psi_1\tilde{p}p-\tilde{s}\phi_1s-\tilde{p}\phi_1p.\ee 
It is now convenient to rewrite the adjoint of $U(2)$ $\phi_1$ as $\eta I_2+\phi$, where $I_2$ is of course the $2\times2$ identity matrix, $\eta = \frac{1}{2}\tr(\phi_1)$ and $\phi$ is the traceless part. Since $\phi^2=\frac{\tr\phi^2}{2}I_2$, we can rewrite the superpotential as 
\be\mathcal{W}=\left(X+\eta^2+\frac{\tr\phi^2}{2}\right)\tilde{q}q+2\eta\tilde{q}\phi q+(\chi-\eta)\tilde{s}s+(\psi_1-\eta)\tilde{p}p-\tilde{s}\phi s-\tilde{p}\phi p.\ee 
By applying chiral ring stability we can simplify the first term and rewrite it simply as $X\tilde{q}q$ since the F-term for $X$ sets $\tilde{q}q$ to zero in the chiral ring. Overall we can rewrite the superpotential as 
\be\label{sup123}\mathcal{W}=X\tilde{q}q+2\eta'\tilde{q}\phi q+\chi'\tilde{s}s+\psi_1'\tilde{p}p-\tilde{s}\phi s-\tilde{p}\phi p,\ee 
where we have also performed the field redefinition $\chi'=\chi-\eta$, $\psi_1'=\psi_1-\eta$ and $\eta' = \eta$. We are then led to the conclusion that $\eta'$ is no longer forced to have the same charges under global symmetries as $\chi',\psi_1'$ and $\phi$ and we gain a new $U(1)$ symmetry under which $X$, $\eta'$ have charge $-2$; $q,\tilde{q}$ have charge 1; and all other fields are uncharged. The issue is that, contrary to the abelian case discussed before, the field redefinition 
\be
(\chi-\eta, \psi_1-\eta, \eta) ~\longrightarrow~ (\chi', \psi', \eta')
\ee
we have just performed is not unitary and makes the K\"{a}hler potential off-diagonal. The requirement that it is uncharged under all global symmetries of the theory reinforces the constraint that $\chi,\eta$ and $\psi_1$ have the same charge. We thus conclude that classically the Lagrangian is not invariant under the aforementioned symmetry. 

Of course, this does not rule out the possibility that it emerges in the infrared. Assuming it does, can we match it with a global symmetry of SQCD? In order to answer this question, we recall that the monopole operators of this theory are mapped to meson components of $SU(2)$ SQCD with three flavors and in particular all the monopole operators with charge $\pm1$ under the topological symmetry of the $U(2)$ central node (whose charge under the aforementioned $U(1)$ symmetry is $\frac{1}{2}[(-2)+(-2)+1+1]=-1$) can be mapped in to meson components of the form $\widetilde{Q}_1Q^i$ and $\widetilde{Q}_iQ^1$ ($i=2,3$).  All other monopole operators are uncharged. Moreover, the operator $qs\tilde{s}\tilde{q}$ which is mapped to the monopole of the $SU(2)$ theory has charge +2. This is precisely compatible with the $U(1)$ symmetry of SQCD which assigns charge $-1$ to $\widetilde{Q}_1$ and $Q_1$ and zero to the other flavors. In other words, this emergent $U(1)$ global symmetry is mapped to a Cartan element of the axial symmetry of SQCD under mirror symmetry. This gives supporting evidence for this emergent symmetry and, moreover, it indicates that the emergent $U(1)$ symmetry does not mix with the $R$-symmetry.  Hence, the emergence of this $U(1)$ global symmetry does not affect the $R$-charge assignments that we use to match the partition functions in section \ref{sec:matchZ}.

Assuming this extra $U(1)$ is there, we find in theory (\ref{sup123}) a rank five global symmetry, coming from three $U(1)$ topological symmetries, one $U(1)$ flavour symmetry and the aforementioned $U(1)$, whereas $SU(2)$ SQCD with three flavors is known to have $U(6)$ symmetry, so we are missing a $U(1)$ generator which is not manifest from the above Lagrangian description. In the case of SQCD with $N$ flavors the global symmetry has rank $2N$, whereas on the dual side we see manifestly $N+2$ $U(1)$ symmetries, including the emergent one. As we have said in the main body of the paper, we leave the discussion of the remaining hidden symmetries for future work. The main difference with respect to the abelian case discussed at the beginning is that the presence of the adjoint chiral multiplet $\phi$ prevents us from assigning independent charges to $\chi'$ and $\psi_1'$. The discussion for $U(2)$ SQCD is unchanged since the superpotential is the same, the only difference being that a $U(1)$ tail is now ungauged. Again, the emergent symmetry in the mirror theory can be matched with the $U(1)$ symmetry acting on one flavor only: in this case we have two monopole operators of charge $\pm1$ under the topological symmetry, which are mapped in the mirror to $qs$ and $\tilde{s}\tilde{q}$ respectively. Both monopole operators of $U(2)$ SQCD have charge +1 under such a symmetry. Again, the SQCD model has a rank six global symmetry whereas in our dual description we see a rank five symmetry group. The mismatch grows linearly with the number of flavors. Following the same reasoning, in the case of $SU(N)$ SQCD we find that the superpotential can be written as follows: 
\be\mathcal{W}=\sum_{i=0}^{N-1}X_i\tilde{q}\phi^iq+\sum_j\alpha_j\tilde{b}_jb^j+\chi'\tilde{s}s+\psi_1\tilde{p}p+\dots\ee 
where $\phi$ denotes again the traceless part of the $U(N)$ adjoint. Every bifundamental is coupled to a different singlet (this is analogous to the abelian case) and the dots stand for cubic terms involving the traceless part of the adjoint chirals $\phi_i$. These are the same as in the $\mathcal{N}=4$ theory.

\section{Quivers with alternating orthogonal and symplectic gauge groups} \label{sec:orthosymp}
In this appendix, we state various conjectures about mirror theories of 3d $\CN=2$ SQCD with orthosymplectic gauge groups and zero superpotential.  The proposed mirror theories involve quivers with alternating orthogonal and symplectic gauge groups. In order to motivate such conjectures, we start with $\CN=4$ mirror pairs of linear quivers.  These models are studied in detail in \cite{Feng:2000eq,Gaiotto:2008ak,Cremonesi:2014kwa, Cremonesi:2014uva, Cabrera:2017njm} and they admit brane realizations.  We then proceed in a similar way as described in section \ref{sec:branes}, namely turn on the nilpotent VEVs for one of the flavour symmetry in the $\CN=4$ mirror theory. In this way, we can obtain the mirror theories of $\CN=2$ SQCD as well as their superpotentials. 

We emphasise that the results in this appendix are conjectural for the following reasons. First of all, we do not have a solid statement of the duality analogous to \eref{dualconfine} for the orthosymplectic gauge group.  One of the reasons is that for the an orthosymplectic gauge group, there is no $U(1)$ topological symmetry and the symmetry generators are usually hidden \cite{Kapustin:1998fa, Feng:2000eq, Cremonesi:2014kwa}.  This makes the explicit charge assignment in the level of Lagrangians difficult. Moreover, as pointed out in \cite{Gaiotto:2008ak}, $\CN=4$ mirror theories of certain linear quivers in this section are ``bad theories'' in the sense that the dimension of some monopole operators falls below the unitarity bound.  In the latter case, the best we could do is to map the ``dressed'' monopole operators in the mirror theory whose dimensions stay above the unitarity bound to the chiral operator of original theory. In any case, since the results could be interesting and potentially be useful for future work, we simply state the results without derivations, along with the $R$-charge of the chiral fields and basic operator maps.

\subsection{$USp(2k)$ gauge theory with $N_f$ flavours}

\subsubsection{$\CN=4$ mirror pairs}
Let us first consider $3d$ $\CN=4$ $USp(2k)$ gauge theory with $N_f$ flavours,
\be \label{USp2NwNf}
\rnode{}{2k}-\sqbnode{}{2N_f}
\ee  
There are two known mirror theories of \eref{USp2NwNf}.  One can be obtained by using the brane construction involving an O3-plane (see \cite[Fig. 13]{Feng:2000eq}):
{\footnotesize
\be \label{mirrUSp2NwNfA}
\bnode{}{2} - \rnode{}{2}-\bnode{}{4} - \rnode{}{4} - \cdots - \bnode{}{2k-2} - \rnode{}{2k-2}  - \bnode{}{2k}  -\rnode{\bluesqver{}{1}}{2k} -  \underbrace{\bnode{}{2k+1}- \rnode{}{2k} - \cdots- \bnode{}{2k+1}}_{\substack{k_f-2k-1\,\,\text{blue nodes} \\ k_f-2k-2\,\,\text{red nodes}}}- \rnode{\bluesqver{}{1}}{2k}    - \bnode{}{2k} - \rnode{}{2k-2}- \bnode{}{2k-2}- \cdots -\bnode{}{4} - \rnode{}{4} -\bnode{}{2} - \rnode{}{2}
\ee}
The other mirror theory can be obtained by using the brane construction involving an O5-plane (see \cite[sec. 4.1.1 \& Fig. 12]{Hanany:1999sj}):
\be \label{mirrUSp2NwNfB}
 \node{}{1}- \node{}{2}-\cdots -\node{}{2k-1}-\underbrace{\node{\wver{}{\,\,1}}{2k} - \node{}{2k} - \ldots - \node{\ver{}{k}}{2k}}_{(k_f-2k-1)~\text{$(2k)$-nodes}}-\node{}{k}~, 
\ee

Observe that we can recover the theory $(B')$ in \eref{u1su2andmirr} from \eref{mirrUSp2NwNfB} as follows. First, we consider $USp(2)$ gauge theory (\ie~$N=1$) with $N_f+1$ flavours:
\be
\sqbnode{}{2}-\rnode{}{2}-\sqbnode{}{2N_f}
\ee
Gauging the $SO(2)$ flavour symmetry in the above quiver, we obtain
\be \label{SO2USp2Nf}
\bnode{}{2}-\rnode{}{2}-\sqbnode{}{2N_f}
\ee
On the mirror side, this amounts to ungauging the leftmost $U(1)$ node (with $k=1$ and $N_f \rightarrow N_f+1$) in \eref{mirrUSp2NwNfB} and hence we obtain
\be \label{mirrSO2USp2Nf}
\underbrace{\node{\wver{}{\,\,2}}{2} - \node{}{2} - \ldots - \node{\ver{}{1}}{2}}_{(N_f-2)~\text{$(2)$-nodes}}-\node{}{1}~, 
\ee
Observe that \eref{SO2USp2Nf} and \eref{mirrSO2USp2Nf} are indeed the mirror pairs in \eref{u1su2andmirr}.  

In addition to \eref{mirrSO2USp2Nf}, one can indeed obtain another mirror theory of \eref{SO2USp2Nf} in a similar manner from \eref{mirrUSp2NwNfA}.  Taking $k=1$ and $N_f \rightarrow N_f+1$ in \eref{mirrUSp2NwNfA} and ungauging the leftmost $SO(2)$ gauge group, we obtain
\be \label{mirrSO2USp2NfB}
\rnode{\bluesqver{}{3}}{2} -  \underbrace{\bnode{}{3}- \rnode{}{2} - \cdots- \bnode{}{3}}_{\substack{N_f-2\,\,\text{blue nodes} \\ N_f-3\,\,\text{red nodes}}}- \rnode{\bluesqver{}{1}}{2}    - \bnode{}{2}
\ee

\paragraph{Generalisation.}  Let us generalise such mirror pairs by considering the following quiver:
\be \label{USp2NwNfa}
(A): \qquad \bnode{}{2}-\rnode{}{2}-\bnode{}{4}-\rnode{}{4} - \cdots -\bnode{}{2k}-\rnode{}{2k}  -\sqbnode{}{2N_f}
\ee  
This theory is also known as $T_{[2N_f-2k-1,1^{2k+1}]}(SO(2N_f))$ in the notation of \cite{Gaiotto:2008ak}.   The mirror of \eref{USp2NwNfa} is denoted by $T^{[2N_f-2k-1,1^{2k+1}]}(SO(2N_f))$.  It admits the following quiver description \cite{Cremonesi:2014uva}:
\be
(B): \qquad \sqbnode{}{2k+1}-\rnode{}{2k}-\underbrace{\bnode{}{2k+1} - \rnode{}{2k}- \cdots-\bnode{}{2k+1}}_{\substack{ N_f-k-1 \, \text{blue nodes} \\ N_f-k-2 \, \text{red nodes}}}-\rnode{\bluesqver{}{1}}{2k}-\bnode{}{2k}- \cdots-\rnode{}{4} -\bnode{}{4}-\rnode{}{2}-\bnode{}{2}
\ee
For $k=1$, this is in agreements with \eref{mirrSO2USp2NfB}.

\subsubsection{$\CN=2$ $USp(2k)$ SQCD with $W=0$ and its mirror}
We obtain the following $3d$ $\CN=2$ mirror pair as in the previous sections:
\be
\begin{split}
(a): &\quad \rnode{}{2k} \overset{Q}{-} \sqnode{}{2N_f}  \quad \text{with $W_{(a')} =0$} \\
(b): &\quad    \bsqnode{}{1} \overset{q}{-}  \rnode{\overset{\phi_1}{\bigcap}}{2k} \overset{b_1}{-}  \underbrace{\bnode{\overset{\phi_2}{\bigcap}}{2k+1} \overset{b_2}{-}  \rnode{\overset{\phi_3}{\bigcap}}{2k} \overset{b_3}{-}  \cdots -  \bnode{\overset{\phi_{m'}}{\bigcap}}{2k+1}}_{\substack{N_f-k-1\,\,\text{blue nodes} \\ N_f-k-2\,\,\text{red nodes}}} \overset{b_{m'}}{-} \underset{\underset{\phi_{m'+1}}{\bigcup}}{\rnode{{}_{s} \! \bsqcver{}{1}}{2k}}   \overset{p_{2k-1}}{-} \bnode{\overset{\psi_{2k-1}}{\bigcap}}{2k}    \cdots \overset{p_4}{-} \rnode{\overset{\psi_4}{\bigcap}}{4}  \overset{p_3}{-} \bnode{\overset{\psi_3}{\bigcap}}{4}   \overset{p_2}{-} \rnode{\overset{\psi_2}{\bigcap}}{2}   \overset{p_1}{-} \bnode{\overset{\psi_1}{\bigcap}}{2} \\
&  \quad \text{with $W_{(b')}$ and $m' = 2N_f-2k-2$.} 
\end{split}
\ee
where the above quivers are written in the $\CN=2$ notation, in which 
\bi
\item each node denotes a 3d $\CN=2$ vector multiplet;
\item each $-$ denotes a chiral multiplet in the $SO \times USp$ bi-fundamental representation; and
\item each $\bigcap$ denotes the adjoint chiral field.
\ei
The superpotential $W_{(b)}$ contains the following terms
\be
q (\phi^{2k+1}_1) q +\sum_{j=0}^{k-1} X_{2j} q \phi_1^{2j} q + \tilde{W}_{N=4} ~,
\ee
where the power of $\phi_1$ in these terms are fixed using the principal orbit $[2k+1]$ of $SO(2k+1)$. Note that the number of flipping fields is equal to the number of independent Casimirs of $USp(2k)$. As before, $ \tilde{W}_{N=4} $ denotes a collection of the cubic superpotential terms that comes from $\CN=4$ supersymmetry. 

The special case of $k=1$ deserves a special attention.
\be \label{mirrpairUSp2}
\begin{split}
(a'): &\qquad \rnode{}{2} \overset{Q}{-} \sqnode{}{2N_f}  \quad \text{with $W_{(a')} =0$} \\
(b'): &\qquad  \bsqnode{}{O(1)} \overset{q}{-}  \rnode{\overset{\phi_1}{\bigcap}}{2} \overset{b_1}{-}  \underbrace{\bnode{\overset{\phi_2}{\bigcap}}{3} \overset{b_2}{-}  \rnode{\overset{\phi_3}{\bigcap}}{2} \overset{b_3}{-}  \cdots -  \bnode{\overset{\phi_{m'}}{\bigcap}}{3}}_{\substack{N_f-2\,\,\text{blue nodes} \\ N_f-3\,\,\text{red nodes}}} \overset{b_{m'}}{-} \underset{\underset{\phi_{m'+1}}{\bigcup}}{\rnode{{}_{s} \! \bsqcver{}{1}}{2}}   \overset{p}{-} \bnode{\overset{\psi_1}{\bigcap}}{2} \\
& \qquad \text{with $W_{(b')}$ and $m'=2N_f-4$.}
\end{split}
\ee
This provides the another duality frame for the $SU(2)$ gauge theory with $N_f$ flavours in addition to \eref{mirrpairSU2}.

\subsubsection*{$R$-charges and operator maps}
Let us denote the $R$-charge of $Q$ in theory $(a)$ by $r$:
\be
R[Q] = r~.
\ee
From the superpotential terms $\phi_{2m+1} b_{2m+1} b_{2m+1}$,  the $R$-charges $R[\phi]$ of $\phi_1$, $\phi_3$, $\phi_5$, $\ldots$, $\phi_{m'-1}$ can be written as
\be \label{RphiUSp}
R[\phi] := R[\phi_{2m+1}] = 2- 2R[b]~,
\ee
where $R[b] :=R[b_1]=R[b_2]= \cdots = R[b_{m'}]$.   

We propose that the meson $M= QQ$ in theory $(a)$ is mapped to the minimal monopole operator $Y^{(b)}$ of any $USp(2k)$ gauge group in theory $(b)$:
\be
M ~\longleftrightarrow ~Y^{(b)}~.
\ee
It follows that
\be
2 r = 2 (2 k + 1) (1 - R[b]) + (2 k) (1 - R[\phi] ) - 2 k~,
\ee
where the right hand side is the $R$-charge of the monopole operator $Y^{(b)}$; see \eg~ \cite[(3.7)]{Amariti:2015vwa}.  Plugging \eref{RphiUSp} into the above equation, we obtain
\be
R[b] = 1-r ~,
\ee
and hence
\be
R[\phi] = 2r~.
\ee

The superpotential term $(\phi^{2k+1}_1) q q$ implies that
\be
2 = 2R[q] +(2k+1) R[\phi]
\ee
We thus obtain the $R$-charge of $q$ to be
\be
R[q] = R[b](1+2k)-2k = (1-r)(1+2k) -2k = 1-(1+2k)r~.
\ee
The $R$-charges of the flipping fields $X_{2j}$ (with $j=0,1,2,\ldots, k-1$) are thus
\be
R[X_j] = 2 - 2 R[q]-2j R[\phi] = 2(1+2k-2j)r~.
\ee

The $R$-charge of the minimal monopole operator $Y^{(a)}$ of the $USp(2k)$ gauge group in theory $(a)$ is
\be
\begin{split}
R[Y^{(a)}] = 2N_f(1-r) -2k~.
\end{split}
\ee
This turns out to be equal to the $R$-charge of the gauge invariant combination $q b_1 b_2 \ldots b_{m'} s$  in theory $(b)$:
\be
R[q b_1 b_2 \ldots b_{m'} s] = R[q] + m' R[b] +R[b] = 2N_f(1-r) -2k~.
\ee
We thus conclude that the minimal monopole operator $Y^{(a)}$ in theory $(a)$ is mapped to the operator in theory $(b)$ as follows:
\be
Y^{(a)} \quad \longleftrightarrow \quad q b_1 b_2 \ldots b_{m'} s~.
\ee
This is as expected from mirror symmetry.

\subsection{$SO(2k)$ gauge theory with $2N_f$ flavours}
\subsubsection{$\CN=4$ mirror pairs}
Let us start by considering the following 3d $\CN=4$ theory:
\be \label{O2NwNf}
(A): \quad \bnode{}{2}-\rnode{}{2}-\bnode{}{4}-\rnode{}{4} - \cdots -\bnode{}{2k-2}-\rnode{}{2k-2}-\bnode{}{2k} -\sqrnode{}{2N_f}
\ee  
where the blue node with a label $m$ denotes an $SO(m)$ group and the red node with an even label $m$ denotes a $USp(m)$ group.  This theory is also known as $T^{[1^{2N_f}]}_{[2N_f-2k+1,1^{2k}]}(USp(2N_f))$ in the notation of \cite{Gaiotto:2008ak}.   The mirror of \eref{O2NwNf} is denoted by $T^{[2N_f-2k+1,1^{2k}]}_{[1^{2N_f}]}(SO(2N_f+1))$, whose quiver is given by \cite{Cremonesi:2014uva}
\be \label{N4mirrSO2k}
(B): \quad \sqbnode{}{2k}-\underbrace{\rnode{}{2k}-\bnode{}{2k} - \rnode{}{2k}- \cdots-\bnode{}{2k}}_{\substack{ N_f-k \, \text{blue nodes} \\ N_f-k \, \text{red nodes}}}-\rnode{\bluesqver{}{1}}{2k}-\bnode{}{2k-1}- \cdots-\rnode{}{4} -\bnode{}{3}-\rnode{}{2}-\bnode{}{1}
\ee

\subsubsection{$SO(2k)$ SQCD with $2N_f$ flavours, $W=0$ and its mirror}
We obtain the following $3d$ $\CN=2$ mirror pair as in the previous sections:
\be
\begin{split}
(a): &\quad \bnode{}{2k} \overset{Q}{-} \sqnode{}{2N_f}  \quad \text{with $W_{(a)} =0$} \\
(b): &\quad    \sqgrnode{}{1} \alr{q}  \rnode{\overset{\phi_1}{\bigcap}}{2k} \overset{b_1}{-}  \underbrace{\bnode{\overset{\phi_2}{\bigcap}}{2k} \overset{b_2}{-}  \rnode{\overset{\phi_3}{\bigcap}}{2k} \overset{b_3}{-}  \cdots -  \bnode{\overset{\phi_{m'}}{\bigcap}}{2k}}_{\substack{N_f-k\,\,\text{blue nodes} \\ N_f-k-1\,\,\text{red nodes}}} \overset{b_{m'}}{-} \underset{\underset{\phi_{m'+1}}{\bigcup}}{\rnode{{}_{s} \! \bluesqver{}{1}}{2k}}   \overset{p_{2k-1}}{-} \bnode{\overset{\psi_{2k-1}}{\bigcap}}{2k-1}    \cdots \overset{p_4}{-} \rnode{\overset{\psi_4}{\bigcap}}{4}  \overset{p_3}{-} \bnode{\overset{\psi_3}{\bigcap}}{3}   \overset{p_2}{-} \rnode{\overset{\psi_2}{\bigcap}}{2}   \overset{p_1}{-} \bnode{\overset{\psi_1}{\bigcap}}{1} \\
&  \quad \text{with $W_{(b)}$ and $m' = 2N_f-2k$.} 
\end{split}
\ee
The superpotential $W_{(b)}$ contains the following terms
\be
\tilde{q} ( \phi_1^{2k-1})  \tilde{q} +q  \phi_1  q + \sum_{j=0}^{k-1} X_{2j} \tilde{q} \phi_1^{2j} \tilde{q} + X_{k} \tilde{q} \phi_1^k \tilde{q} + \tilde{W}_{N=4} ~,
\ee
where the power of $\phi_1$ in these terms are fixed using the principal orbit $[2k-1,1]$ of $SO(2k)$ in the same way as in \cite{Maruyoshi:2016tqk}. Note that the number of flipping fields is equal to the number of independent Casimirs of $SO(2k)$.

\subsubsection*{$R$-charges and operator maps}
Let us denote the $R$-charge of $Q$ in theory $(a)$ by $r$:
\be
R[Q] = r~.
\ee
From the superpotential terms $\phi_{2m+1} b_{2m+1} b_{2m+1}$,  the $R$-charges $R[\phi]$ of $\phi_1$, $\phi_3$, $\phi_5$, $\ldots$, $\phi_{m'-1}$ can be written as
\be \label{RphiSOeven}
R[\phi] := R[\phi_{2m+1}] = 2- 2R[b]~,
\ee
where $R[b] :=R[b_1]=R[b_2]= \cdots = R[b_{m'}]$.   

Let us denote the monopole operator in any $USp(2k)$ gauge group in theory $(b)$ dressed with the adjoint matter field $\phi$ by
\be
Y^{(b)}_j = \tr(Y^{(b)}_0 \phi^j)
\ee
where $Y^{(b)}_0$ is the minimal monopole operator of any $USp(2k)$ gauge group in theory $(b)$.   The $R$-charge of $Y_j^{(b)}$ is  (see \eg~ \cite[(3.7)]{Amariti:2015vwa}):
\be
R[Y_j^{(b)}] = 2 (2 k) (1 - R[b]) + (2 k) (1 - R[\phi]) + j R[\phi] - 2 k \overset{\eref{RphiSOeven}}{=} 2j (1- R[b])~.
\ee
Let us point out that for $j=0$, $R[Y_0^{(b)}] =0$.  This means that the dimension of the minimal monopole operator $Y_0^{(b)}$ falls below the unitary bound.   Indeed for the theory with $\CN=4$ supersymmetry, namely \eref{N4mirrSO2k} with $R[b]=1/2$, a $USp(2k)$ gauge group with $2k$ flavours renders the theory ``bad'' in the sense of \cite{Gaiotto:2008ak}.  Hence, to make sense of this, we consider $Y^{(b)}_j$ with $j \geq 1$.

We propose that the meson $M= QQ$ in theory $(a)$ is mapped to the monopole operator $Y_1^{(b)}$ in theory $(b)$:
\be
M ~\longleftrightarrow ~Y_{1}^{(b)}~.
\ee
It follows that
\be
2 r = R[Y^{(b)}_1 ] = 2(1-R[b])~.
\ee
Thus,
\be
R[b] = 1-r ~.
\ee
We therefore obtain
\be
R[\phi] = 2r~.
\ee
The $R$-charge of the operator $Y_j^{(b)}$ is thus
\be
R[Y_j^{(b)}] = 2jr~.
\ee

The superpotential term $\phi_1 q q$ imposes the condition
\be
R[q] = R[b] = 1-r~.
\ee
The superpotential term $( \phi_1^{2k-1}) \tilde{q} \tilde{q}$ imposes the condition
\be
2 R[\tilde{q}] + (2 k - 1) R[\phi] = 2~,
\ee
and so
\be
 R[\tilde{q}] = 2  + 2 k R[b] - R[b] - 2 k = 1-(2k-1)r~.
\ee

The $R$-charges of the flipping fields are as follows:
\bea
R[X_{2j}] &= 2 - 2R[\tilde{q}]-2jR[\phi] = 2(2k-1-2j) r~, \quad j=0,\ldots, k-1 \\
R[X_k] &= 2-2R[\tilde{q}]-kR[\phi]=2r (k-1)~.
\eea

The minimal monopole operator $Y^{(a)}$ of gauge group $SO(2k)$ theory $(a)$ has $R$-charge
\be
R[Y^{(a)}] = 2 N_f (1 - r) - (2 k - 2)~.
\ee
The baryon-monopole operator $\beta^{(a)}$ of theory $(a)$ has $R$-charge
\be
R[\beta^{(a)}] =(2k-2) r+R[Y^{(a)}] = 2 N_f (1 - r) - (2 k - 2)(1-r)~.
\ee
These can be matched with the following $R$-charges of the operators of $(b)$:
\be
\begin{split}
R[q b_1 b_2 \ldots b_{m'} s] &=R[q] +m' R[b] + R[b] = 2 N_f (1 - r) - (2 k - 2)(1-r) \\
R[s b_{m'} \ldots b_2 b_1 \tilde{q}] &= R[b] + m'R[b] +R[\tilde{q}] = 2 N_f (1 - r) - (2 k - 2) ~.
\end{split}
\ee
We thus propose that $Y^{(a)}$ and $\beta^{(a)}$ are mapped to the operators of $(b)$ as follows:
\be
\begin{split}
&\beta^{(a)} \quad \longleftrightarrow \quad q b_1 b_2 \ldots b_{m'} s \\
&Y^{(a)} \quad \longleftrightarrow \quad s b_{m'} \ldots b_2 b_1 \tilde{q} 
\end{split}
\ee
Moreover, the $R$-charge of the baryon $B$ in theory $(a)$ is
\be
R[B] = 2kr~.
\ee
The baryon $B$ in theory $(a)$ is mapped to the monopole operator in theory $(b)$ as follows:
\be
B \quad \longleftrightarrow \quad Y_k^{(b)}~.
\ee


\subsection{$O(2k+1)$ gauge theory with $2N_f$ flavours}
\subsubsection{$\CN=4$ mirror pairs}
Let us start by considering the following 3d $\CN=4$ theory:
\be \label{OoddwNf}
\bnode{}{1}-\rnode{}{2}-\bnode{}{3}-\rnode{}{4} - \cdots -\rnode{}{2k}-\bnode{}{2k+1} -\sqrnode{}{2N_f}
\ee  
This theory is also known as $T^{[1^{2N_f}]}_{[2N_f-2k,1^{2k}]}(USp(2N_f)')$ in the notation of \cite{Cremonesi:2014uva}.   The mirror of \eref{OoddwNf} is denoted by $T^{[2N_f-2k,1^{2k}]}_{[1^{2N_f}]}(USp(2N_f)')$, whose quiver is given by
\be
\sqrnode{}{2k}-\underbrace{\bnode{}{2k+2}-\rnode{}{2k} - \bnode{}{2k+2}-\rnode{}{2k}- \cdots-\bnode{}{2k+2}}_{\substack{ N_f-k \, \text{blue nodes} \\ N_f-k-1 \, \text{red nodes}}}-\rnode{\bluesqver{}{1}}{2k}-\bnode{}{2k+1}- \cdots-\rnode{}{4} -\bnode{}{5}-\rnode{}{2}-\bnode{}{3}
\ee

\subsubsection{$O(2k+1)$ SQCD with $2N_f$ flavours, $W=0$ and its mirror}
We obtain the following $3d$ $\CN=2$ mirror pair as in the previous sections:
\be \label{N4mirrSO2kp1}
\begin{split}
(a): &\quad \bnode{}{O(2k+1)} \overset{Q}{-} \sqnode{}{2N_f}  \quad \text{with $W_{(a)} =0$} \\
(b): &\quad    \sqgrnode{}{SU(1)} \overset{q}{-}  \bnode{\overset{\phi_1}{\bigcap}}{2k+2} \overset{b_1}{-}  \underbrace{\rnode{\overset{\phi_2}{\bigcap}}{2k} \overset{b_2}{-}  \bnode{\overset{\phi_3}{\bigcap}}{2k+2} \overset{b_3}{-}  \cdots -  \bnode{\overset{\phi_{m'}}{\bigcap}}{2k+2}}_{\substack{N_f-k-1\,\,\text{blue nodes} \\ N_f-k-1\,\,\text{red nodes}}} \overset{b_{m'}}{-} \underset{\underset{\phi_{m'+1}}{\bigcup}}{\rnode{{}_{s} \! \bsqcver{}{1}}{2k}}   \overset{p_{2k-1}}{-} \bnode{\overset{\psi_{2k-1}}{\bigcap}}{2k+1}    \cdots \overset{p_4}{-} \rnode{\overset{\psi_4}{\bigcap}}{4}  \overset{p_3}{-} \bnode{\overset{\psi_3}{\bigcap}}{5}   \overset{p_2}{-} \rnode{\overset{\psi_2}{\bigcap}}{2}   \overset{p_1}{-} \bnode{\overset{\psi_1}{\bigcap}}{3} \\
&  \quad \text{with $W_{(b)}$ and $m' = 2N_f-2k-1$.} 
\end{split}
\ee
The superpotential $W_{(b)}$ contains the term
\be
q(\phi_1^{2k})q + \sum_{j=0}^{k} X_{2j}q \phi_1^{2j} q + \tilde{W}_{N=4}~.
\ee
%
%

\subsubsection*{$R$-charges and operator maps}
Let us denote the $R$-charge of $Q$ in theory $(a)$ by $r$:
\be
R[Q] = r~.
\ee
From the superpotential terms $\phi_{2m+1} b_{2m+1} b_{2m+1}$,  the $R$-charges $R[\phi]$ of $\phi_1$, $\phi_3$, $\phi_5$, $\ldots$, $\phi_{m'-1}$ can be written as
\be \label{RphiSOodd}
R[\phi] := R[\phi_{2m+1}] = 2- 2R[b]~,
\ee
where $R[b] :=R[b_1]=R[b_2]= \cdots = R[b_{m'}]=R[s]$.   

Let us denote the monopole operator in any $SO(2k+2)$ gauge group in theory $(b)$ dressed with the adjoint matter field $\phi$ by
\be
Y^{(b)}_j = \tr(Y^{(b)}_0 \phi^j)
\ee
where $Y^{(b)}_0$ is the minimal monopole operator of any $SO(2k+2)$ gauge group in theory $(b)$.   The $R$-charge of $Y_j^{(b)}$ is  (see \eg~ \cite[(3.7)]{Amariti:2015vwa}):
\be
\begin{split}
R[Y_j^{(b)}] &= 2 (2 k) (1 - R[b]) + [(2 k+2)-2] (1 - R[\phi]) + j R[\phi] \\ & \qquad - [(2 k+2)-2] \\
&\overset{\eref{RphiSOodd}}{=} 2(1- R[b])(j-k)~.
\end{split}
\ee
Let us point out that for $0 \leq j \leq k$, $R[Y_0^{(b)}]\leq 0 $, assuming that $0 \leq R[b] <1$.  Indeed for the theory with $\CN=4$ supersymmetry, namely \eref{N4mirrSO2kp1} with $R[b] =1/2$, a $SO(2k+2)$ gauge group with $2k$ flavours renders the theory ``bad'' in the sense of \cite{Gaiotto:2008ak}.  Hence, to make sense of this, we consider $Y^{(b)}_j$ with $j \geq k+1$.

We propose that the meson $M= QQ$ in theory $(a)$ is mapped to the monopole operator $Y_{k+1}^{(b)}$ in theory $(b)$:
\be
M ~\longleftrightarrow ~Y_{k+1}^{(b)}~.
\ee
It follows that
\be
2 r = R[Y^{(b)}_{k+1} ] = 2(1-R[b])~.
\ee
Thus,
\be
R[b] = 1-r ~,
\ee
and hence
\be
R[\phi]=2r~.
\ee
The $R$-charge of the operator $Y_j^{(b)}$ is thus
\be
R[Y_j^{(b)}] = 2r(j-k)~.
\ee

The superpotential term $\phi_1^{2k} q q$ imposes the condition
\be
2R[q]+2k R[\phi] =2~.
\ee
Therefore,
\be
R[q] = 1- k R[\phi]= 1-2kr~.
\ee
The $R$-charges of the flipping fields $X_{2j}$ (with $j=0,1, \ldots, k$) are thus
\be
R[X_{2j}] = 2- 2R[q]-2jR[\phi]=4r (k-j)~.
\ee

The minimal monopole operator $Y^{(a)}$ of gauge group $SO(2k)$ theory $(a)$ has $R$-charge
\be
R[Y^{(a)}] = 2 N_f (1 - r) - (2 k+1 - 2)~.
\ee
This can be matched with the following $R$-charges of the operators of $(b)$:
\be
\begin{split}
R[q b_1 b_2 \ldots b_{m'} s] &=R[q] +m' R[b] + R[b] = 2 N_f (1 - r) - (2 k+1 - 2)~.
\end{split}
\ee
We thus propose that $Y^{(a)}$ is mapped to the operator of theory $(b)$ as follows:
\be
\begin{split}
&Y^{(a)} \quad \longleftrightarrow \quad q b_1 b_2 \ldots b_{m'} s~.
\end{split}
\ee

\bibliographystyle{ytphys}
\bibliography{ref,ref1}

\end{document}